\documentclass[fleqn,usenatbib]{mnras}

\usepackage{newtxtext,newtxmath}

\usepackage[T1]{fontenc}
\usepackage{ae,aecompl}

\usepackage{graphicx}	
\usepackage{amsmath}	
\usepackage{amsbsy}	
\usepackage{array}		
\usepackage{gensymb}	
\usepackage{tikz}
\usepackage{makecell}

\defcitealias{VTST:2000}{VTST00}

\title[Non-Relativistic MHD jet models]{A study of radial self-similar non-relativistic MHD outflow models: parameter space exploration and application to the water fountain W43A}

\author[Ceccobello, C. et al.]{
Ceccobello, C.$^{1}$\thanks{E-mail: chiara.ceccobello@chalmers.se}
Heemskerk, M.H.M.$^{2}$, Cavecchi, Y.$^3$, Vlemmings, W.H.T.$^{1}$, Tafoya, D.$^{1}$ 
\\
$^{1}$Department of Space, Earth and Environment, Chalmers University of Technology, Onsala Space Observatory, 439 92 Onsala, Sweden\\
$^{2}$Anton Pannekoek Institute for Astronomy, University of Amsterdam, Science Park 904, 1098 XH Amsterdam, The Netherlands\\
$^{3}$ Mathematical Sciences and STAG Research Centre, University of Southampton, SO17 1BJ, UK
}

\date{Accepted XXX. Received YYY; in original form ZZZ}

\pubyear{2020}

\begin{document}
\label{firstpage}
\pagerange{\pageref{firstpage}--\pageref{lastpage}}
\maketitle

\begin{abstract}
Outflows, spanning a wide range of dynamical properties and spatial extensions, have now been associated with a variety of accreting astrophysical objects, from supermassive black holes at the core of active galaxies to young stellar objects.
The role of such outflows is key to the evolution of the system that generates them, for they extract a fraction of the orbiting material and angular momentum from the region close to the central object and release them in the surroundings. The details of the launching mechanism and their impact on the environment are fundamental to understand the evolution of individual sources and the similarities between different types of outflow-launching systems.
We solve semi-analytically the non-relativistic, ideal, magnetohydrodynamics (MHD) equations describing outflows launched from a rotating disk threaded with magnetic fields using our new numerical scheme. 
We present here a parameter study of a large sample of new solutions. We study the different combinations of forces that lead to a successfully launched jet and discuss their global properties. 
We show how these solutions can be applied to the outflow of the water fountain W43A for which we have observational constraints on magnetic field, density and velocity of the flow at the location of two symmetrical water maser emitting regions.
\end{abstract}




\section{Introduction}
\label{sec:intro}
 Jets, and more generally speaking outflows, are a widespread phenomena in many different systems, from protostars to supermassive black holes. 
 Until only recently, outflows were believed to be launched by extraction of rotational energy either from a magnetically-threaded disk \citep{BlandfordPayne:1982} or from a rotating black hole \citep{BlandfordZnajek:1977}. In fact, it was common to think that there were at least two separated kinds of jets, the magnetically-dominated jets from black holes and the pressure-dominated jets from any other jetted source. 
However, with the advent of cutting-edge GRMHD simulations and the first post-processed  emission spectrum associated with it \citep{Moscibrodzka:2016,Liska:2017, Davelaar:2019}, it is becoming clear that there is not such a dichotomy and, most likely, at least in black hole systems, the two mechanisms can coexist. Furthermore, the emission is likely dominated by the outer, more mass loaded, jet sheath rooted onto the accretion disk, whereas the inner core of the jet is lighter and magnetically-dominated \citep{Moscibrodzka:2016}.
Similarly, when the jet-launching object is a protostar or a (non-BH) compact object, the outflow is likely to be a composition of a stellar wind \citep[e.g.][]{Shu:1994} or an equivalent Blandford-Znajek process for highly magnetized neutron stars  \citep{Parfrey:2016} and a disk-driven outflow (e.g. \citealp{PudritzNorman:1983,ContopoulosLovelace:1994,Ferreira:1997,VTST:2000}). 

Semi-analytical models describing the various launching mechanisms listed above have been continuously developed in parallel with simulations because they capture the underlying physics while allowing a time-efficient exploration of the parameter space and fitting of astrophysical sources. 
However, in order to make the equations treatable with a semi-analytical approach, the dimensionality of the problem is reduced by assuming symmetries in the system and a non-linear separation of variables is performed.
The separation of variables is commonly referred to as the \emph{self-similarity} assumption. There are two distinct classes of self-similar models depending on how the separation of variables is carried out \citep{VlahakisTsinganos:1998}: the \emph{meridional} self-similar models \citep[e.g.][]{Sauty:1994,Trussoni:1997,Sauty:1999,Chantry:2018}, where the dependent variables are functions of $r$ and \emph{radial} self-similar models \citep[e.g.][]{ContopoulosLovelace:1994,Ferreira:1997,VTST:2000, VlahakisKonigl:2003, Polko:2010,Polko:2013,Polko:2014,Ceccobello:2018}, where the independent variable is $\theta$.
In both classes of self-similar models we are left with a mixed system of differential and algebraic equations describing the accelerating flow along a magnetic field line threading a rotating disk. To determine the motion of the fluid element, one needs to solve simultaneously the forces acting along and perpendicular to a streamline.  
This is a notoriously cumbersome problem, which can be tackled by introducing further simplifications, such as assuming a fixed structure of the magnetic field and/or neglecting the gas pressure force and/or using asymptotic extensions of the models to replace the region of the solutions at large distance from the disk.

In \citet[][hereafter Paper I]{Ceccobello:2018}, we presented our newly developed algorithm to self-consistently solve the poloidal and transverse forces, given by the Bernoulli and Grad-Shafranov equations respectively, for a relativistic fluid in the presence of gravity, under the assumption of radial self-similarity. We showed that with our numerical algorithm it is possible to obtain solutions with a broad variety of jet structures and dynamical properties and work is ongoing to couple these solutions with a radiative code and apply those to black hole systems (Lucchini et al. submitted).
In this paper, we adopt the equations presented in \citet[VTST00 from now on]{VTST:2000} and we adapt our algorithm, described in Paper I, to perform a parameter study to model astrophysical sources with more moderate speeds, such as young stellar objects (YSOs) and evolved stars outflows.

In Sec.~\ref{sec:eqnMet} we summarize the basic equations and give a short description of the algorithm. In Sec.~\ref{sec:parstu}, we show the results of our parameter space exploration and discuss the solution properties as they transition from cold jets to hot ones. In Sec.~\ref{sec:app}, we show an example of an application to the post-AGB star W43A and we give the selection criteria we used to isolate the solutions that better resemble the jet of W43A and discuss the characteristics of the selected jet configuration in relation to the source.
Finally in Sec.~\ref{sec:discussion}, we summarise the study presented in this paper.


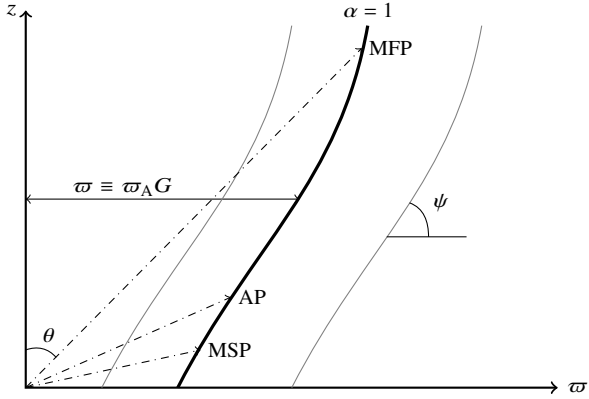
\begin{figure}
    \centering
    \begin{tikzpicture}[scale=1]
    \draw [thick, black,<->] (0,5) node[left] {$z$} --(0,0) -- (7,0) node[right]  {$\varpi$};
    \draw[<->] (0,2.5) node[right,xshift=0.5cm,yshift=0.16cm] {$\varpi\equiv \varpi_{\rm A} G$} -- (3.6,2.5);
    \draw[ very thick] (2,0) to [out=63,in=-100] (4.5,4.8) node[black,above] {$\alpha=1$};
    \draw[gray] (3.5,0) to [out=63,in=-100] (6.0,4.8);
     \draw[gray] (1.0,0) to [out=63,in=-100] (3.5,4.8);
    \draw[dashdotted,->] (0,0) -- (2.3,0.5) node[right] {MSP};
    \draw[dashdotted,->] (0,0) -- (2.7,1.2) node[right] {AP};
    \draw[dashdotted,->] (0,0) -- (4.42,4.5) node[right] {MFP};
    \draw (0,0.5) node[above,xshift=0.3cm] {$\theta$} to [out=20,in=130] (0.4,0.4);
   \draw (5.05,2.45) node[right,xshift=0.2cm] {$\psi$} to [out=-20,in=90]  (5.30,2);
   \draw (4.75,2) -- (5.8,2);
    \end{tikzpicture}\hfill
           \caption{System of coordinates we adopt to describe a solution of the MHD system of equations presented in VTST. We identify a solution with the "reference" streamline identified with the label $\alpha \equiv \varpi_{\rm A}^2/\varpi_*^2 = 1$. The two dependent variables $(M,G)$, together with all the other quantities describing the system, are functions of $\theta$ (the independent variable in radial self-similarity), which is the angle between a point on the streamline and the $z$-axis. The angle $\psi$ is defined by the tangent to the streamline and the horizontal axis, while the distance from a point on the streamline to the $z$-axis is defined by its cylindrical radius $\varpi$ in units of the Alfv\'en cylindrical radius $\varpi_{\rm A}$.}
  \label{fig:angles}
   \end{figure}

\section{Equations and numerical method}
\label{sec:eqnMet}

\subsection{Problem description}
\label{ssec:problem}

The equations that we are going to solve with our numerical algorithm are the ones describing an axisymmetric, radial self-similar, non-relativistic, disk-driven outflow with non-negligible enthalpy (Fig.~\ref{fig:angles}). Since we adopted the prescription given in \citet{VTST:2000}\footnote{We will use $\Gamma$ for the polytropic index  and $F$ for the power law exponent, instead of the symbols $\gamma$ and $x$ as was done in \citetalias{VTST:2000} to maintain the same convention we had for the relativistic equations in Paper I.}, we present here just a brief summary. In Appendix \ref{app:scaling} we report the conversion from dimensionless to physical quantities as a function of the input parameters and the scaling relations. The dependent variables of the equations described in \citetalias{VTST:2000} are the poloidal Mach number $M$, the dimensionless cylindrical radius, $G$, and the angle describing the inclination of the streamline with respect to the disk plane, $\psi$. These are all functions of $\theta$ once their radial dependence has been defined as power laws of the function $\alpha \equiv \varpi_{\rm A}^2/\varpi_*^2$, where $\varpi_{\rm A}$ is the cylindrical radius at the Alfv\'en point and $\varpi_*$ is the chosen scaling length of the problem and effectively is the cylindrical radius of the Alfv\'en point on the streamline with $\alpha=1$. (see Fig.~\ref{fig:angles}). 

To obtain a full solution, i.e. a streamline rooted at the disk midplane and terminating infinity, the adopted numerical scheme must handle three singular points that are present in the Bernoulli and Grad-Shafranov equations when solved simultaneously: the Alfv\'en point (AP) and the magnetosonic fast/slow points (MFP/MSP)\footnote{Note that after the separation of variables, they are points (not surfaces) on a single streamline and they are \emph{modified} because their position and definition of the phase speeds of the slow and fast magnetosonic waves are affected by the geometry of the magnetic field \cite[e.g.][]{Sauty:1994,FerreiraPelletier:1995}}.
At each singular point the equations can be regularized either analytically, in the case of the AP, or numerically, for the MSP and the MFP, analogously to the simpler case of the sonic point in the Parker wind model \citep{Parker:1958}. 
The AP has been studied extensively, due to the possibility of manipulating the equations analytically there.
The other two singular points present a more complex case. On the one hand, the position of both the MSP and the MFP is not known before the full solution for a given set of initial parameters is calculated, on the other hand, a full solution cannot be computed without knowing the position of these two singular points and the AP. 
Due to this intrinsic difficulty, the MSP and MFP are often neglected by assuming cold flows, i.e. thermal pressure plays no role in accelerating the flow (no MSP), and/or by adopting a given asymptotic behaviour of the streamline once the flow has become superalfvenic, which effectively pushes the MFP at infinity. Typically, either one or both of the above assumptions are made to avoid dealing with the complexity of determining these singular points.
Moreover, when the MSP and/or the MFP are not removed from the equations, finding solutions across large volumes of the parameter space is a difficult task that requires a solid numerical algorithm capable of recovering the unknown positions of the singular points and properly handling the equations at these locations for wide ranges of the input parameters.
However, the role of the MFP in self-similar theories is fundamental when solving the Bernoulli and Grad-Shafranov equations combined, because it is the singular point where the flow loses causal contact with the source \citep{LCB:1992,Bogovalov:1999,Meier:2012}. 
Downstream of the MFP the flow starts to focus rapidly towards the polar axis up until the last recollimation point (LRP). We identify the LRP with the region where the jet terminates in our solutions (see Paper I).
This region has been connected in relativistic jets with the standing shock/particle acceleration regions in active galactic nuclei and in stellar-mass black hole systems \citep[e.g.][]{Ceccobello:2018,Cohen:2014, Meier:2012,Polko:2010,Markoff:2001,Markoff:2005,Markoff:2010}. 

\citet{WeberDavis:1967} showed that there can be multiple families of solutions with different velocity profiles, crossing either none or one/two/three singular points. We are looking at those that cross all three points, which are characterised by an increasing poloidal Mach number.
\citetalias{VTST:2000} were the first to calculate complete solutions with all these characteristics for the non-relativistic case.

\subsection{Non-relativistic MHD system of equations}
\label{ssec:equations}

The Bernoulli and Grad-Shafranov equations for a steady-state axisymmetric system describe the energy flux balance along the poloidal direction and the equilibrium configuration of the magnetic field lines. 

Both can be derived from the conservation of momentum equation, which describes the forces acting on a streamline:
\begin{equation}
\rho(\boldsymbol{V}\cdot \nabla) \boldsymbol{V} -\frac{1}{4\pi}\boldsymbol{B} \times (\nabla\times \boldsymbol{B}) + \nabla P -\rho \nabla \frac{\mathcal GM}{r} =0
\label{eq:consMom}
\end{equation}
where $\rho, P, \boldsymbol{V}, \boldsymbol{B}$ are the density, pressure, velocity and magnetic field of the flow. $\mathcal{G}$ and $\mathcal{M}$ are the gravitational constant and the mass of the central object, respectively. 

If we adopt either cylindrical ($\boldsymbol{\hat z}, \boldsymbol{\hat \varpi}, \boldsymbol{\hat \phi}$) or spherical coordinates ($\boldsymbol{\hat r}, \boldsymbol{\hat \theta}, \boldsymbol{\hat \phi}$), the poloidal and perpendicular unit vectors ($\hat{\boldsymbol{b}}$,  $\hat{\boldsymbol{n}}$) can be written as follows
\begin{eqnarray}
&\hat{\boldsymbol{n}}  = \cos(\psi)\boldsymbol{\hat z} - \sin(\psi)  \boldsymbol{\hat \varpi} = \cos(\theta+\psi)\hat{\boldsymbol{r}} - \sin(\theta+\psi)\hat{\boldsymbol{\theta}}\\
&\hat{\boldsymbol{b}} = \sin(\psi)\boldsymbol{\hat z} + \cos(\psi)  \boldsymbol{\hat \varpi} = \sin(\theta+\psi)\hat{\boldsymbol{r}}+ \cos(\theta+\psi)\hat{\boldsymbol{\theta}} \\
&\hat{\boldsymbol{\phi}} = \hat{\boldsymbol{\phi}}.
\end{eqnarray}

The projection of Eq.~\ref{eq:consMom} along $\hat{\boldsymbol{b}}$, the Bernoulli equation, describes how the different types of energies can be converted to one another. The projection of Eq.~\ref{eq:consMom} along $\hat{\boldsymbol{n}}$, the Grad-Shafranov equation or transfield equation, provides the shape of the magnetic field lines. 

The projections of the Bernoulli and the transfield equation can be rewritten using the scaling equations given in Appendix \ref{app:scaling} and then rearranged in the following form:
\begin{equation}
A_i\frac{dM^2}{d\theta} + B_i\frac{d\psi}{d\theta} = C_i
\label{eqn:detform}
\end{equation}
with $i=1$ representing the coefficients of the Bernoulli equation and $i=2$ the coefficients of the transfield equation.
The Bernoulli and transfield equations arranged in the way described above can further be recast into a system of two first-order differential equations for the evolution of the poloidal Mach number $M(\theta) = \sqrt{4\pi \rho} V_p/B_p$ and the angle $\psi$ describing the inclination of the streamline with respect to the horizontal axis:
\begin{align}
\frac{dM^2}{d\theta} & = \frac{\mathcal{N_{\rm 1}}}{\mathcal{D}}  =  \frac{B_2 C_1 - B_1 C_2}{ A_1B_2- A_2 B_1}\label{eqn:dM2dt}\\
\frac{d\psi}{d\theta} & = \frac{\mathcal{N_{\rm 2}}}{\mathcal{D}}  =  \frac{ A_1 C_2 - A_2 C_1}{ A_1B_2- A_2 B_1}, \label{eqn:dpsidt}
\end{align}
with the numerators $\mathcal{N}_i$ ($i=1,2$) and the denominator $\mathcal{D}$ being functions of the coefficients $A_i,B_i,C_i$ ($i=1,2$) which are given in Appendix~\ref{app:numsden}.

As described in paper I, in order to minimize the intrinsic errors we chose not to solve Eq.~\ref{eqn:dpsidt}, but instead derive $\psi(\theta)$ from the Bernoulli integral Eq.~\ref{eqn:BernConst} \citepalias[see also][and Appendix]{VTST:2000} from the MSP to the LRP.
Upstream of the MSP, the streamlines can undergo oscillations, depending on the given set of input parameters, so the sign of $\cos(\psi + \theta)$ can change. Hence Eq.~\ref{eqn:dpsidt} must be integrated with care in this region to ensure the correct radial profile of the solutions from the disk to the MSP.

Additionally, we solve a differential equation for the unknown function $G(\theta)$, which is
defined as the cylindrical radius to the polar axis of a streamline
labeled by $\alpha$, normalised to its cylindrical radius at the
Alfv\'en point. The equation for $G$ is the following
\begin{align}
G(\theta)& = \frac{\varpi}{\varpi_\alpha} = \frac{\varpi}{\varpi_\star} \alpha^{-1/2}\\
\frac{dG^2}{d\theta} &= \frac{2 G^2\cos(\psi)}{\sin(\theta)\cos(\psi+\theta)}.
\label{eqn:dG2dt}
\end{align}
The solution of these equations depends on six parameters: $\Gamma, F, k_{\rm VTST},\lambda_{\rm VTST}, \mu_{\rm VTST}$ and $\epsilon_{\rm VTST}$ \citepalias[see][]{VTST:2000}. 
The first parameter $\Gamma$ is the polytropic index in the equation of state $q = P/\rho^{\Gamma}$, where $q$ is the specific gas entropy and a constant of motion of the problem. The parameter $F$ determines the initial current distribution in the radial direction, $-\varpi B_\phi = \mathcal C_2 \sin(\theta) r^{F-1}$, which is an increasing or decreasing function of $r$ depending on the value of $F$. This parameter also determines the radial dependence of the magnetic field lines through $B\sim \varpi^{F-2}$. $k_{\rm VTST}$ is proportional to the ratio between the Keplerian speed and the poloidal flow speed at the Alfv\'en radial distance, and often is referred to as the \emph{mass loss parameter} (see e.g. \citealp{Ferreira:1997}, but also \citetalias{VTST:2000}). $\lambda_{\rm VTST}$ is the specific angular momentum in units of $V_\star \varpi_\star$ and $\mu_{\rm VTST}$ is proportional to the gas entropy.
The parameters $k_{\rm VTST}, \mu_{\rm VTST}$ and $\lambda_{\rm VTST}$ are defined by the following relations:
\begin{equation}
k_{\rm VTST} = \sqrt{\frac{\mathcal{GM}}{\varpi_\star V_\star^2}}; \quad \mu_{\rm VTST} = \frac {8 \pi P_\star}{B_\star^2}; \quad
\lambda_{\rm VTST} = \frac{L}{V_\star \varpi_\star}.
\end{equation}
It is worth noticing that the starred quantities found across the paper are scaling factors and can be related to the quantities calculated at the AP on the reference streamline ($\alpha=1$), namely $\rho_\star = \rho_{\rm A}$, $\varpi_\star = \varpi_{\rm A}$ and
\begin{equation}
\left(B_*,V_*\right) = -\frac{\cos(\theta_{\rm A} + \psi_{\rm A})}{\sin(\theta_{\rm A})}\left(B_{p,\rm A}, V_{p,\rm A}\right), \quad {\rm with} B_* = \sqrt{4\pi\rho_*} V_*
\end{equation}
Finally, $\epsilon_{\rm VTST}$ is the sum of kinetic, enthalpy, gravitational and Poynting energy flux densities per unit of mass flux density, rescaled by $\alpha^{-1/2}V^{2}_\star$, i.e.
\begin{align}
\epsilon_{\rm VTST} &= \frac{\alpha^{1/2}}{V_\star^2} E =   \left[\epsilon_{\rm{K}, p} + \epsilon_{\rm{K}, \phi} +\epsilon_{\rm{T}} +\epsilon_{\rm{M} }+ \epsilon_{\rm G} \right] \nonumber\\
& =  \left[\frac{1}{2} \left(\frac{M^2}{G^2}\frac{\sin(\theta)}{\cos(\theta+\psi)}\right)^2 + \frac{1}{2}\left(\frac{\lambda_{\rm VTST}}{G^2}\frac{G^2 - M^2}{1-M^2}\right)^2 \right. \nonumber\\
& \left. \quad+ \frac{\mu_{\rm VTST}}{2} \frac{\Gamma}{\Gamma-1}M^{2(1-\Gamma)} + \lambda_{\rm VTST}^2\frac{1-G^2}{1-M^2}\right.\nonumber\\
& \left.\quad - k_{\rm VTST}^2 \frac{\sin(\theta)}{G}\right]. \label{eqn:BernConst}
\end{align}
The total energy flux per unit mass can be rescaled with the Alfv\'en poloidal velocity as
$2 E/V_{\rm A,p}^2 = 2\epsilon_{\rm VTST} V_*^2\alpha^{-1/2}/V_{\rm A,p}^2$, which becomes
\begin{equation}
\tilde \epsilon =  2\epsilon_{\rm VTST} \frac{\cos^2(\theta_{\rm A}+\psi_{\rm A})}{\sin^2(\theta_{\rm A})} 
\end{equation}
and with the use of the De L'H\^opital rule to regularize the indefinite terms ( see Eq.~\ref{eqn:delHop}), we can write it at the AP and obtain the Alfv\'en Regularity Condition \citepalias[ARC, see][]{VTST:2000} in the compact form
\begin{align}
 \tilde{\epsilon} =&   1 + \frac{\cos^2(\theta_{\rm A}+\psi_{\rm A})}{\sin^2(\theta_{\rm A})}\left[- 2k_{\rm VTST}^2 \sin(\theta_{\rm A}) + \mu_{\rm VTST}\frac{\Gamma}{\Gamma-1} \right.\nonumber\\
 & \left. +\lambda_{\rm VTST}^2\left(1+g_{\rm A}^2\right)\right]. 
\label{eqn:ARCrescaled}
\end{align}
The function $g_{\rm A}$ is the \emph{fastness} parameter calculated at the AP. A general definition of the fastness parameter given by \citet{PelletierPudritz:1992} is
\begin{equation}
\frac{V_\phi}{\varpi}  = \Omega (1- g) 
\label{eqn:g}
\end{equation}
where 
\begin{equation}
\Omega = \frac{1}{\varpi} \left(V_\phi - V_p \frac{B_\phi}{B_p}\right) 
\label{eqn:Omega}
\end{equation}
is the angular frequency of the streamline, which is a constant of motion of the problem.
The fastness parameter gives a measure of how large  the angular velocity of the gas is in relation to the angular velocity of the magnetic surface on which it moves.
We can derive $g_{\rm A}$ from the application of the De L'H\^opital rule to the indefinite forms
\begin{align}
\left.\frac{1-G^2}{1-M^2}\right|_{\rm A} & \equiv g_{\rm A}   =  \frac{2 \cos(\psi_{\rm A})}{p_{\rm A} \sin(\theta_{\rm A})\cos(\theta_{\rm A}+\psi_{\rm A})}\\
\left.\frac{G^2-M^2}{1-M^2}\right|_{\rm A}&  = 1 - g_{\rm A}
\label{eqn:delHop}
\end{align}
where $p_{\rm A} = dM^2/d\theta |_{\rm A}$.
In the following section, we summarise the method we developed in Paper I that we now adapt to solve the non-relativistic equations. For the details of the algorithm, we address the interested reader to Paper I. Indeed, there is no substantial difference in the mechanics of the algorithm, although the non-relativistic equations are noticeably easier to handle.

\begin{table}
\setlength\tabcolsep{4.5 pt}
\setlength\extrarowheight{5pt}
 \caption{Model parameters. The parameters are equivalent to \citetalias{VTST:2000}, but we changed the notation of some of them to avoid confusion with the relativistic parameters and physical quantities described in Paper I.}
 \label{tab:modpars}
 \begin{tabular}{l | m{5cm} }
  \hline
 Input parameters &\\
  \hline
  $F$					& exponent of the radial scaling of the current\\
  $\Gamma$ 			& polytropic index of the gas \\
  $\theta_{\rm A}$		& angular distance of the AP from the jet axis\\
  $\psi_{\rm A}$   		&  inclination of the streamline with respect to the horizontal axis at the AP
  \\
  $k_{\rm VTST}$		& 	mass loss parameter \\
  \hline
  Fitted parameters & \\
  \hline
  $\theta_{\rm MFP}$  		& angular distance of the MFP from the jet axis  \\
  $\theta_{\rm MSP}$   		& angular distance of the MSP from the jet axis  \\
  $\mu_{\rm VTST} $ 		& scaling of the gas-to-magnetic pressure ratio\\
  $\lambda_{\rm VTST}$ 		& specific angular momentum in
units of $V_\star \varpi_\star$ \\
  \hline
  \end{tabular}
  \end{table}

\subsection{Method}
\label{ssec:met}

In Paper I, we described a new numerical method to find solutions to the relativistic radial self-similar MHD equations for a disk-launched jet in the presence of gravity \citep{VlahakisKonigl:2003, Polko:2014}. 
As discussed in Sec.\ref{ssec:problem}, even under the simplifying assumption of self-similarity, solving self-consistently and simultaneously the Bernoulli and Grad-Shafranov equations is known to be a rather difficult task because of the singular surfaces. 
At the location of the singular points, the equations \ref{eqn:dM2dt}-\ref{eqn:dpsidt} are indeterminate but finite, e.g. 
\begin{equation}
\frac{dM^2}{d\theta} = \frac{\mathcal{N}_1}{\mathcal{D}} =  \frac00 = {\rm finite}. \label{eqn:windSing}
\end{equation}
However, only at the AP one can derive an analytical expression that gives the finite value of the derivative of the poloidal Mach number (Alfv\'en Regularity Condition, ARC). The location of the AP and $G^2_{\rm A}, M^2_{\rm A}, dG^2/d\theta |_{\rm A}$ and $dM^2/d\theta |_{\rm A}$ can be determined from the values of the input parameters and the ARC (Eq.~\ref{eqn:ARCrescaled}).
The regularity conditions at the MFP and MSP can exclusively be derived numerically together with their position on the streamline. 

As a result, the most frequent approach is to determine all the unknown functions and parameters at AP and then integrate the system with a shooting method towards the other two singular points. However, given the high accuracy needed to determine the values of the parameters and the intrinsic numerical difficulties of treating, under these conditions, the form 0/0, this method presents serious drawbacks and does not allow to easily find and convincingly identify solutions to the required accuracy threshold. Therefore, it impedes a full exploration of the parameter space.

The structure of our numerical method is the following:
\begin{itemize}
\item[1.] We guess the locations of the critical points,  $\theta_{\rm MSP}$ and $\theta_{\rm MFP}$, and derive values for $M^2, G^2$ and their derivatives given by the condition that the numerators and the denominator of  Eq.~\ref{eqn:windSing}, and of the similar equation for $\psi$, i.e. $d\psi/d\theta = \mathcal{N}_2/\mathcal{D} = 0/0$, are zero at the MSP/MFP of choice.
\item[2.] We integrate away from AP, MSP and MFP towards the midpoints $\theta_{\rm mid, MSP} = (\theta_{\rm A} + \theta_{\rm MSP})/2$ and $\theta_{\rm mid, MFP} = (\theta_{\rm A} + \theta_{\rm MFP})/2$
\item[3.] We determine the parameters that give a match at the midpoints using the Bayesian open-source code \texttt{multinest} \citep{Multinest:2008, Multinest:2009,Multinest:2013}.
\end{itemize}
The specific choice of input parameters and fitted parameters is given in Tab.~\ref{tab:modpars}. Once a particular family of solutions is specified through the choice of $F,\Gamma, \theta_{\rm A},\psi_{\rm A}$and $k_{\rm VTST}$, we identify the location of the MSP and MFP and the best-fit values of the remaining parameters $\mu_{\rm VTST}$ and $\lambda_{\rm VTST}$ and we extend the solutions upstream of the MSP towards the disk midplane and downstream of the MFP towards the last recollimation point (LRP). In paper I, we defined this point as the last point we were able to calculate with our algorithm. The last few integration points before LRP seem to indicate the onset of a recollimation shock where the fluid is compressed in a small section around the polar axis. Indeed, we noticed that the denominator is approaching zero again in Eq.~\ref{eqn:dM2dt} and Eq.~\ref{eqn:dpsidt}, while the numerator is not. This means that both the derivative of $M^2$ and $\psi$ become infinite close to LRP, making the integration towards this (singular) point impossible.

\begin{figure}
\centering
\includegraphics[width=1\linewidth]{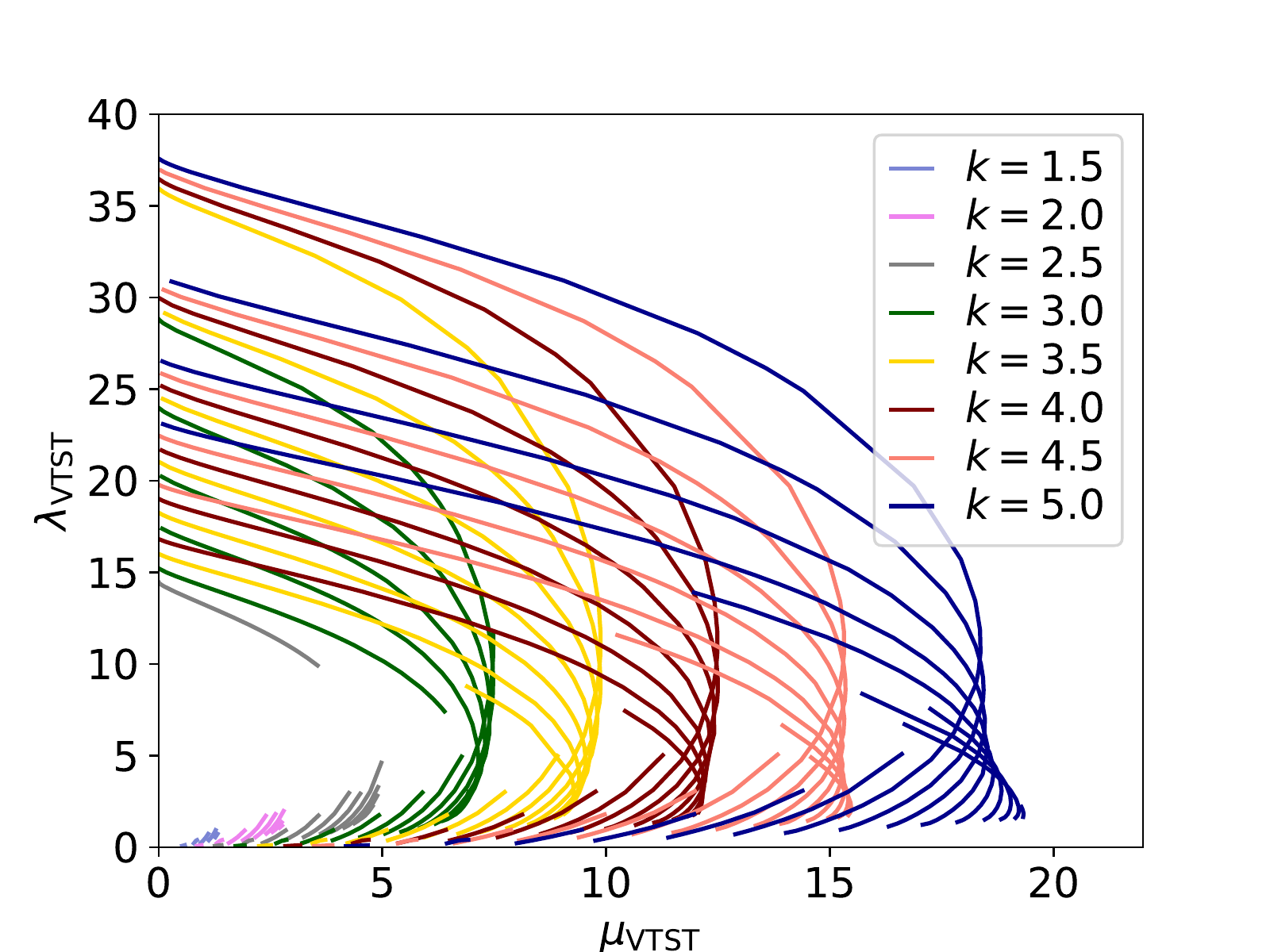}\hfill\\
\includegraphics[width=0.88\linewidth, angle=0]{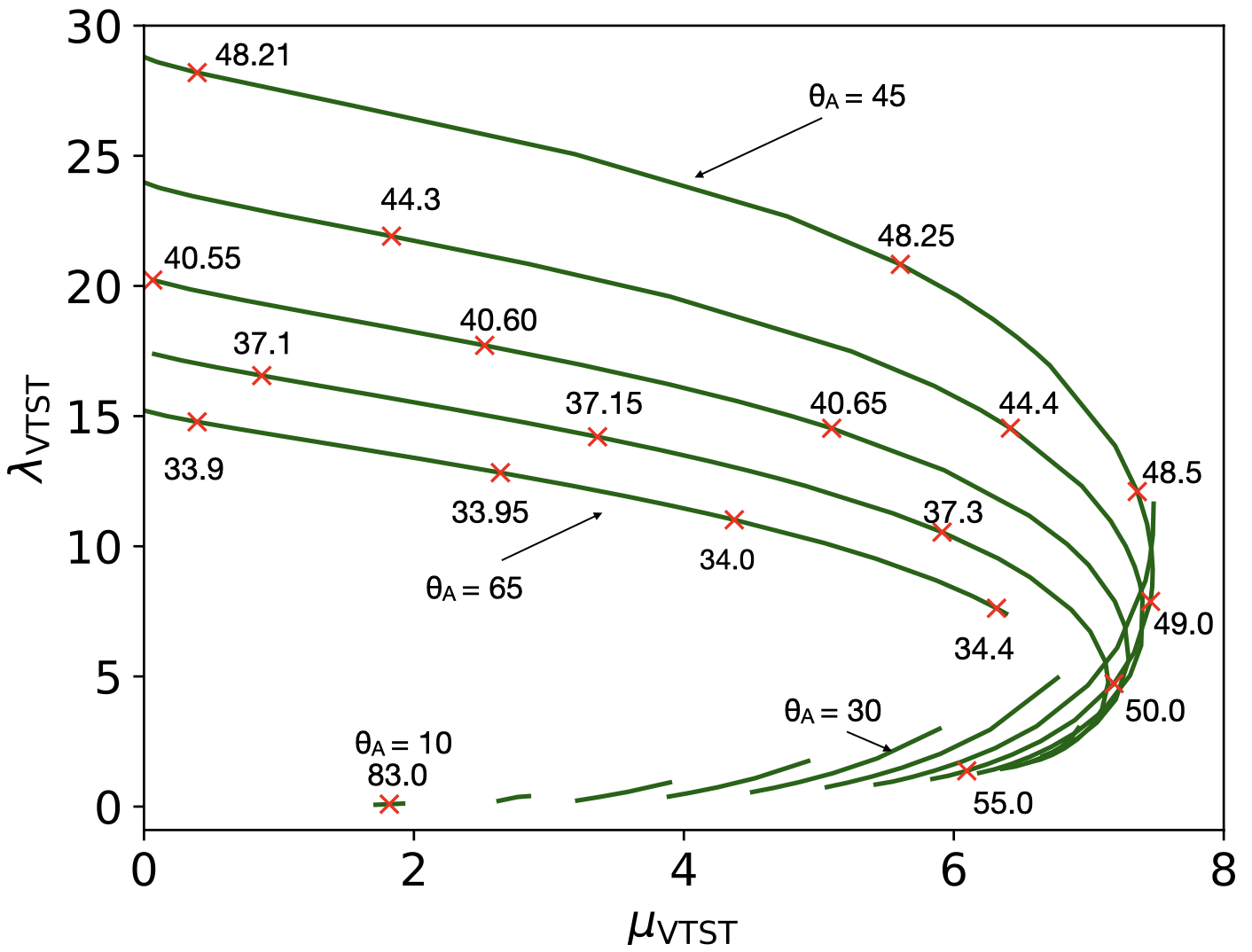}
\caption{\emph{Upper panel:} Grid of solutions presented in the angular momentum and entropy plane, i.e. $(\lambda_{\rm VTST}$,$\mu_{\rm VTST})$-plane for $\Gamma=5/3$, $F=0.75$. The location ($\theta_{\rm A}$) of the AP, collimation ($\psi_{\rm A}$) at the AP,  and the mass loss parameter $k_{\rm VTST}$ are allowed to vary within a chosen grid. In each $k_{\rm VTST}=$ constant subset, the lines connect solutions with constant $\theta_{\rm A}$ and variable $\psi_{\rm A}$. \emph{Lower panel:} Same plot for a subset of solutions for $k_{\rm VTST} =3.0$. Neighbouring lines differ in $\theta_{\rm A}$ by 5 degrees. Along each line we have indicated the value of $\psi_{\rm A}$ for some of them to illustrate how sensitive the equations are for a small change of this angle. Particularly, a tiny change in $\psi_{\rm A}$ translates into a large step in $\mu_{\rm VTST}$, when $\lambda_{\rm VTST}$ is large.}
\label{fig:lambdaMu}
\end{figure}

\section{Parameter study}
\label{sec:parstu}

Given the wealth of solutions that we are able to retrieve using this algorithm, we focus on a grid of solutions obtained by fixing the adiabatic index $\Gamma$ to 5/3, the exponent of the radial scaling of the current $F$ to $0.75$, as in \citep[][hereafter BP]{BlandfordPayne:1982}, and the mass loss parameter $k_{\rm VTST}$ between 1.5 and 5.0 in steps of 0.5. We note that the resulting solutions will be generally different from the stereotypical BP-like solution because we include gas pressure and the crossing of all the three singular points.
For each $k_{\rm VTST}$, we seek solutions with all the allowed combinations of $\theta_{\rm A}$ and $\psi_{\rm A}$, which are the angles determining the position and the collimation of the streamline at the Alfv\'en point, respectively. In Fig.~\ref{fig:lambdaMu} we show the distribution of these solutions in the plane of dimensionless angular momentum and entropy, i.e. the $(\lambda_{\rm VTST}$, $\mu_{\rm VTST})$-plane. 
Each line represents solutions for a constant $k_{\rm VTST}$ and $\theta_{\rm A}$, while only $\psi_{\rm A}$ varies.

As the upper panel of Fig.~\ref{fig:lambdaMu} shows, although our solutions cover a good extent of this region of the parameter space, a few series could not be completed because of the disappearing of the MSP below the disk midplane, e.g. for $k_{\rm VTST}=2.5$ (gray line), which are physically not meaningful. 
In the lower panel of Fig.~\ref{fig:lambdaMu} we show how the collimation angle at the AP, $\psi_{\rm A}$, is changing for a few lines on which the position of the AP, $\theta_{\rm A}$, is constant and the mass loss parameter, $k_{\rm VTST}$, has been set to 3.0 for all the lines. We see that the parameters of a solution change significantly with only a small change in  $\psi_{\rm A}$. This is particularly true in the top part of the figure where the dimensionless angular momentum, $\lambda_{\rm VTST}$, is large.

We only find a solution when the sum of the angles $\theta_{\rm A}+ \psi_{\rm A} $ is roughly within the interval $93\degree$-$111\degree$. This range varies depending on the value of  $k_{\rm VTST}$  (see Tab.~\ref{tab:angles}). In general the allowed range of this sum is between $90\degree$ and $180\degree$ to ensure that the derivative of the poloidal Mach number is negative, i.e. the fluid is accelerating at Alfv\'en. For a constant location of the AP, $\theta_{\rm A}$, the collimation angle $ \psi_{\rm A} $ is small when the entropy $\mu_{\rm VTST}$ approaches zero and is large when the angular momentum $\lambda_{\rm VTST}$ approaches zero. As we will discuss later, the combination of these two angles ultimately determines the dynamics and the geometry of the jet and the narrower range of their sum that we find is likely due to the minimum and maximum energy fluxes allowed in this region of the parameter space (see Fig.~\ref{fig:RatioEne}).   

\begin{table}
\centering
\setlength\tabcolsep{4.5 pt}
\setlength\extrarowheight{5pt}
 \caption{The maximum value of $\theta_{\rm A}$ and the sum $\theta_{\rm A}+ \psi_{\rm A} $ with a constant $k_{\rm VTST}$. The minimum of $\theta_{\rm A}=10\degree$ and of the sum  $\theta_{\rm A}+ \psi_{\rm A} =93\degree$ are the same for all $k_{\rm VTST}$ values.}
 \begin{tabular}{l | c c}
 \hline
 $k_{\rm VTST}$ &  $\theta_{\rm A, max}$ &  $(\theta_{\rm A}+ \psi_{\rm A})_{\rm max}$  \\
  \hline
  1.5 & 30\degree &  98\degree  \\
    \hline
  2.0 & 45\degree &  102\degree  \\
    \hline
  2.5 & 65\degree &  106\degree  \\
    \hline
  3.0 & 65\degree &  107\degree  \\
    \hline
  3.5 & 70\degree &  109\degree  \\
    \hline
  4.0 & 75\degree &  110\degree  \\
    \hline
  4.5 & 75\degree &  110\degree  \\
    \hline
  5.0 & 80\degree &  111\degree  \\
    \hline
  \end{tabular}
  \label{tab:angles}
  \end{table}


The lowest value of the sum, i.e. 93 (small $\theta_{\rm A}$, large $\psi_{\rm A}$), coincides with the jet configurations with lowest total energy-to-mass flux ratios which is around $\sim V_{\rm A,p}^2/2$ at the jet base (z=0) for $k_{\rm VTST}$ of the order of unity (Eq.~\ref{eqn:ARCrescaled}). These solutions have little-to-none magnetic field ($\lambda_{\rm VTST} \rightarrow 0$), and represent a tenuous jet (low total energy flux, see Sec.~\ref{ssec:trends} and Sec.~\ref{ssec:hotcold}) supported by some ($\mu_{\rm VTST}$ small) gas pressure, which provides the balance to gravity. 
By varying the two angles within the allowed range, we recover a large collection of solutions where we see low-energy hot jets transform into cold and fast jets with a large angular momentum ($\lambda_{\rm VTST}\gtrsim 20$) and a relatively small contribution of the gas pressure (small $\mu_{\rm VTST}$) to the total energy. 
The large variety of physical properties within this sample of solutions provides an ideal framework to study the different jet configurations and to devise a method for the comparison of such solutions to astrophysical sources.

\subsection{General trends}
\label{ssec:trends}

\begin{figure}
\centering
\includegraphics[width=1\linewidth]{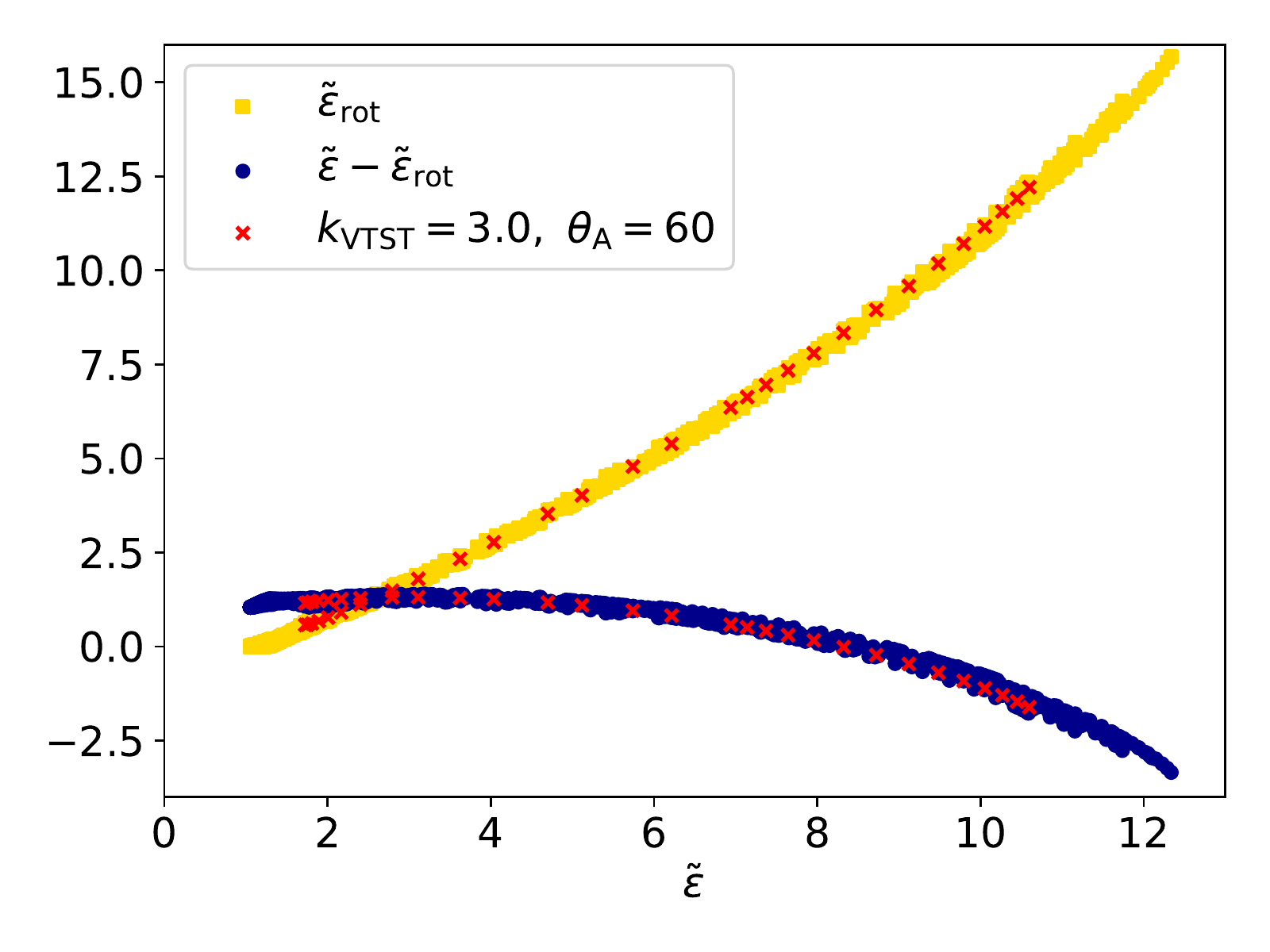}\hfill
\caption{The rotational energy as a function of the total energy scaled with the poloidal velocity at Alfv\'en for all solutions (yellow dots). In blue is shown the total energy in a frame rotating with the constant angular frequency $\Omega$ versus the total energy in the inertial frame, also scaled with the poloidal velocity at Alfv\'en. The red crosses are solutions with $k_{\rm VTST} =3.0$ and $\theta_{\rm A}=60\degree$. They will be used to discuss other trends later on.}
\label{fig:maxEneRotFrame}
\end{figure}

In this Section we discuss some general properties and trends observed while inspecting the whole ensemble of solutions. 
In Fig.~\ref{fig:maxEneRotFrame} we show how the total energy is divided up between rotational energy and generalized pressure \citep{Ferreira:1997}.
The rotational energy is the difference between the total energy in an inertial frame and the total energy in a frame rotating with a frequency $\Omega$ (Eq.~\ref{eqn:Omega}), i.e.
\begin{equation}
E_{\rm rot} = L\Omega = \varpi_{\rm A}^2\Omega^2.
\end{equation}
In the above equation, $L$ is the angular momentum, defined as 
\begin{equation}
L = \varpi \left(V_\phi - \frac{B_\phi}{4\pi \rho}\frac{B_p}{V_p} \right),
\label{eqn:L}
\end{equation}
which is also a constant of motion along the streamline.
Fig.~\ref{fig:maxEneRotFrame} shows the total energy in the rotating frame rescaled by the poloidal kinetic energy at the AP (blue dots):
\begin{equation}
2\frac{E - L\Omega}{V_{\rm A,p}^2} = \tilde \epsilon - \tilde \epsilon_{\rm rot}
\end{equation}
and the rotational energy rescaled, $2L\Omega/V_{\rm A,p}^2= \tilde\epsilon_{\rm rot}$ (yellow dots) versus the rescaled total energy in the inertial frame $\tilde\epsilon = 2E/V_{{\rm A},p}^2$. 

All the points lie on two narrow curves. The solutions highlighted in the bottom panel of Fig.~\ref{fig:lambdaMu} are marked as red crosses in Fig.~\ref{fig:maxEneRotFrame}. The total energy in the rotational frame, $\tilde\epsilon - \tilde\epsilon_{\rm rot}$ (blue dots), otherwise called the generalized pressure \citep{Ferreira:1997,PelletierPudritz:1992}, achieves a maximum when. the rotational energy, $\tilde \epsilon_{\rm rot}$ (yellow dots), is negligible. Since the  total energy flux sustaining a jet, i.e. the Bernoulli constant, is positive, the generalized pressure can change sign depending on the relative contribution of the rotational energy, $\tilde \epsilon_{\rm rot}$, to the total energy.
As the rotational energy, $\tilde \epsilon_{\rm rot}$, increases, it approaches equipartition with the generalized pressure which occurs in the regime where the latter is still positive. When the sign flip occurs, we start to see a dominant contribution of the magnetic energy in the Bernoulli equation (Eq.~\ref{eqn:BernConst}).

\begin{figure}
\centering
\includegraphics[width=1\linewidth]{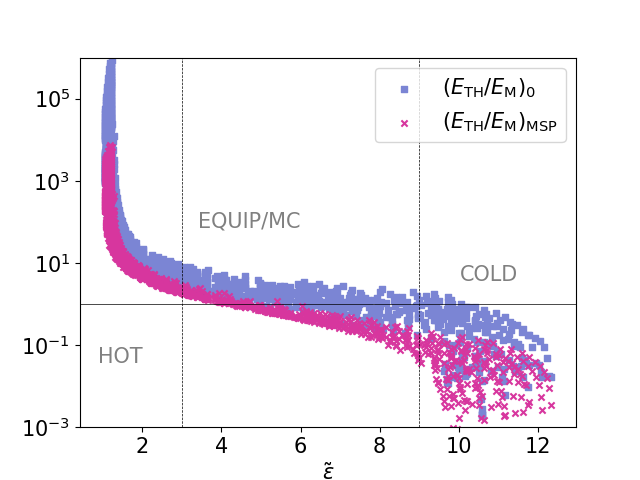}\hfill
\caption{
Ratio between the thermal energy and magnetic energy as a function of the total energy with respect to the poloidal kinetic energy at Alfv\'en for all solutions.}
\label{fig:RatioEne}
\end{figure}

Based on the ratio between the thermal energy and the magnetic energy flux we distinguish three categories of solutions: thermally-dominated hot, equipartition/centrifugal and magnetically-dominated cold jets (see Fig.~\ref{fig:RatioEne}). We show this ratio at the disk midplane (blue squares) and at the MSP (pink crosses) for the full sample of solutions versus the total energy rescaled with the poloidal kinetic energy. The vertical lines are drawn to guide the eye.
We see that the distribution of the jet models in this plane is very similar between $z=0$ and the MSP.
The hot jets are low-energy solutions and as the $E_{\rm TH}/E_{\rm M}$ increases the total energy flux, $\tilde \epsilon$, remains constant and at its minimum value. 
When the thermal and magnetic energy flux are roughly at equipartition, the total energy is increasing steadily as the solutions become more magnetically-dominated. As we enter the cold regime, the magnetic energy grows more rapidly for a small variation of the input parameters (see bottom panel of Fig.~\ref{fig:lambdaMu}), but the jet configurations do not increase so much in total energy anymore, approaching its maximum. Since there is this correspondence between total energy and hot/cold regime, we will use it interchangeably across the paper.

\begin{figure}
\centering
\includegraphics[width=1\linewidth]{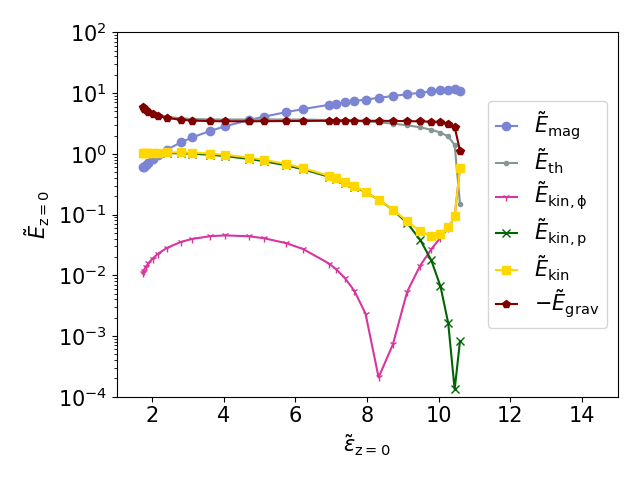}\\
\includegraphics[width=1\linewidth]{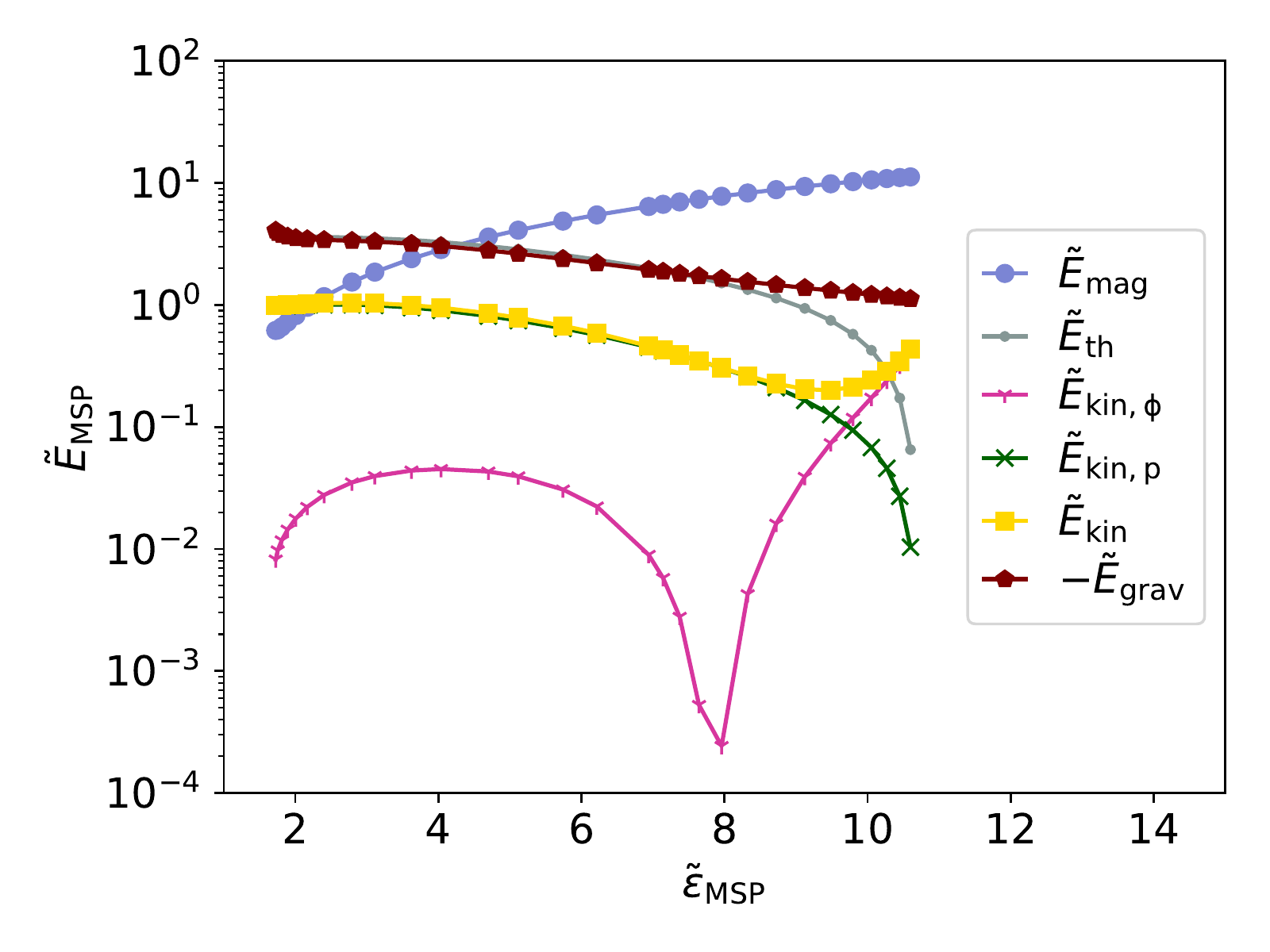}\hfill
\caption{The total energy at the base (top panel) and at the MSP (bottom panel) split up in its components for a series of solutions for $k_{\rm VTST} =3.0$ and $\theta_{\rm A}=60\degree$. All energies have been scaled with the poloidal kinetic energy at Alfv\'en.}
\label{fig:EneCompsSeries}
\end{figure}

In Fig.~\ref{fig:EneCompsSeries} we show the different contributions to the total energy at the base (top panel) and at the MSP (bottom panel) for a series of solutions with $k_{\rm VTST} =3.0$ and $\theta_{\rm A}=60\degree$. The trends discussed here are also observed in other series. We start by noticing that when the magnetic energy is larger, the total energy is larger too. When the total energy is low, the gravitational energy and the thermal energy dominate with almost equal magnitude, cancelling each other. Only at higher total energies the thermal energy becomes negligible. Apart from the most energetic solutions, the kinetic energy consists mainly of the poloidal component. At higher energies the poloidal component of the velocity of the gas leaving the midplane is relatively low while the toroidal speed gives the largest contribution to the  total kinetic energy. 

\begin{figure}
\centering
\includegraphics[width=1\linewidth]{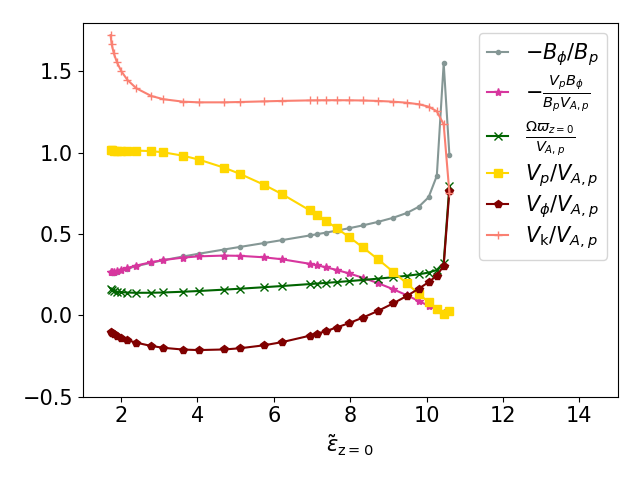}\\
\includegraphics[width=1\linewidth]{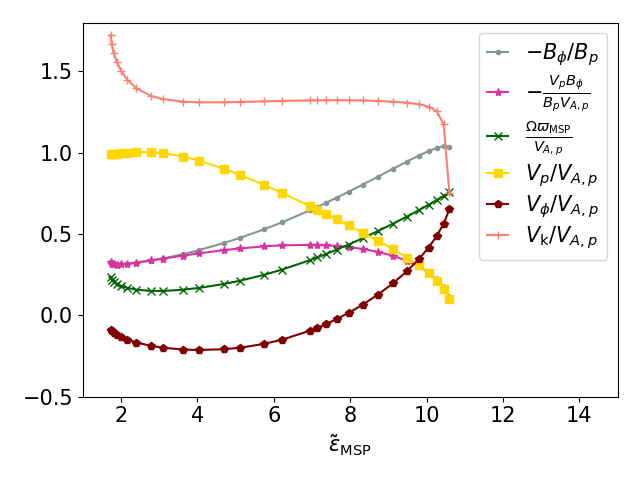}\hfill
\caption{Velocities at the base (top panel) and at the MSP (bottom panel) scaled by the poloidal velocity at Alfv\'en as a function of the scaled total energy for a series of solutions for $k_{\rm VTST} =3.0$ and $\theta_{\rm A}=60\degree$.}
\label{fig:VelCompSeries}
\end{figure}

In Fig.~\ref{fig:VelCompSeries} we present the components of the velocity and of the angular frequency $\Omega$ of the streamlines (Eq.~\ref{eqn:Omega}) for the same series of solutions presented in Fig.~\ref{fig:EneCompsSeries}. At lower total energies the poloidal velocity is relatively large with respect to the toroidal velocity. As a consequence, even when the ratio of the magnetic field components (gray line with dots) is low, i.e. the magnetic field is almost not twisted at all, the second term on the rhs of the equation describing $\Omega$ (Eq.~\ref{eqn:Omega}, magenta line with stars) is dominant, while the toroidal velocity (brown line with pentagons) is negative and smaller. This means that the gas is lagging behind the rotation of the disk and the magnetic field is weak, while $\Omega$ is at its minimum.

Only the very last solution with the highest energy of this series is rotating at keplerian speed, which can be seen by noticing that the last dot of the pink line with crosses ($V_{\rm k}$)
coincides with the last point of the brown line with pentagons ($V_\phi$)
in the top panel of Fig.~\ref{fig:VelCompSeries}.
As expected by the non-negligible contribution of the enthalpy, the overwhelming majority of the solutions in this ensemble is subkeplerian at the disk midplane, with a deviation increasingly larger as the solutions become warmer and warmer.

This means that the typical approximation $\Omega\sim\Omega_{\rm gas} = V_\phi/\varpi  \sim \Omega_{\rm k}$ cannot be taken as a general property of this sample of solutions. Only a small fraction of the solutions presented in this paper can be considered corotating with the disk, like for instance the last three high-energy solutions in Fig.~\ref{fig:VelCompSeries}, where we see that $\Omega\varpi$ (green line with crosses) matches $V_\phi$ (brown line with pentagons), while $-V_{p}B_\phi/B_p$ (magenta line with stars) is close to zero. 

\begin{figure}
\centering
\includegraphics[width=1\linewidth]{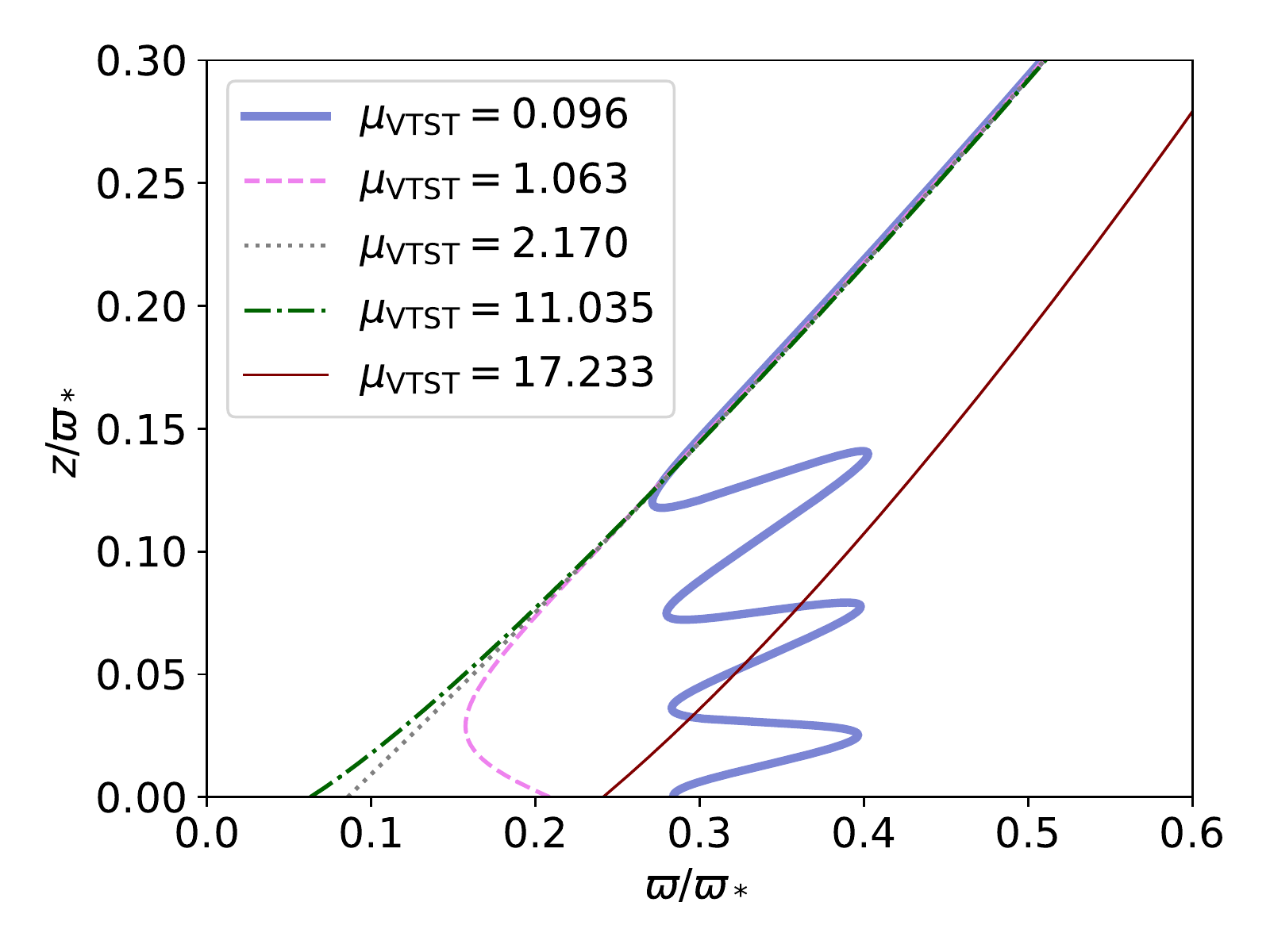}\hfill
\caption{Examples of streamlines of a series of solutions for increasing $\mu_{\rm VTST}$}
\label{fig:streamlines}
\end{figure}

From a geometrical point of view, the radial profile of the streamlines varies depending on how hot the jet is, typically with highly oscillating jet bases for cold jets while no oscillations are present for warm and hot jets (see Fig.~\ref{fig:streamlines}). This is a consequence of the oscillatory nature of the transverse component of the forces that define the collimation of the streamline. We will discuss this topic in detail in Section~\ref{ssec:hotcold}. 
Since it is very likely that such oscillations may be unstable and considering that the MSP is a more robust point in our solutions, we identify the MSP with the jet base from now on.
 
\begin{figure}
\centering
\includegraphics[width=1\linewidth]{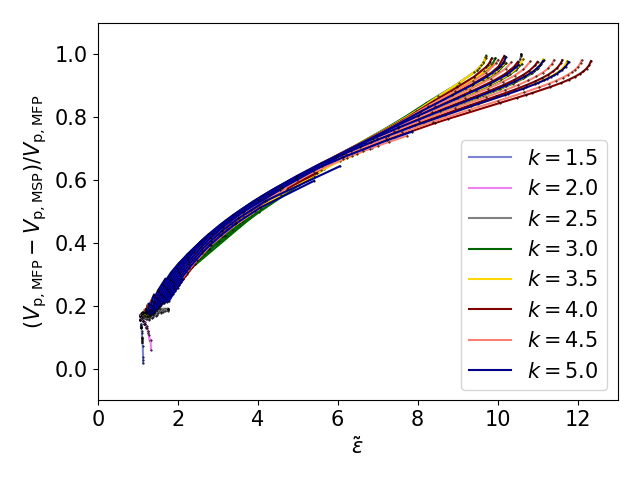}\hfill
\caption{Acceleration fraction versus the total energy rescaled.}
\label{fig:deltaV}
\end{figure}

Different jet configurations can be also classified based on the amount of acceleration that the gas experiences from the MSP to the MFP, being that the point where the flow loses causal contact with the source and the flow upstream. In Fig.~\ref{fig:deltaV}, we plot all the solutions divided in subgroups with constant $k_{\rm VTST}$ in the plane defined by the increase in the poloidal velocity experienced by the matter from the MSP to the MFP and the rescaled total energy flux. 
The low-energy flux, pressure-driven, solutions have also low $\Delta V/V_{\rm MFP}$ since they are characterised by large poloidal velocities at the MSP which do not increases much approaching the MFP. 
As the energy flux increases, the poloidal velocity decreases (see bottom panel of Fig. \ref{fig:VelCompSeries}) and the increment of the velocity $\Delta V/V_{\rm p, MFP}$ approaches 1.

\begin{figure}
\centering
\includegraphics[width=1\linewidth]{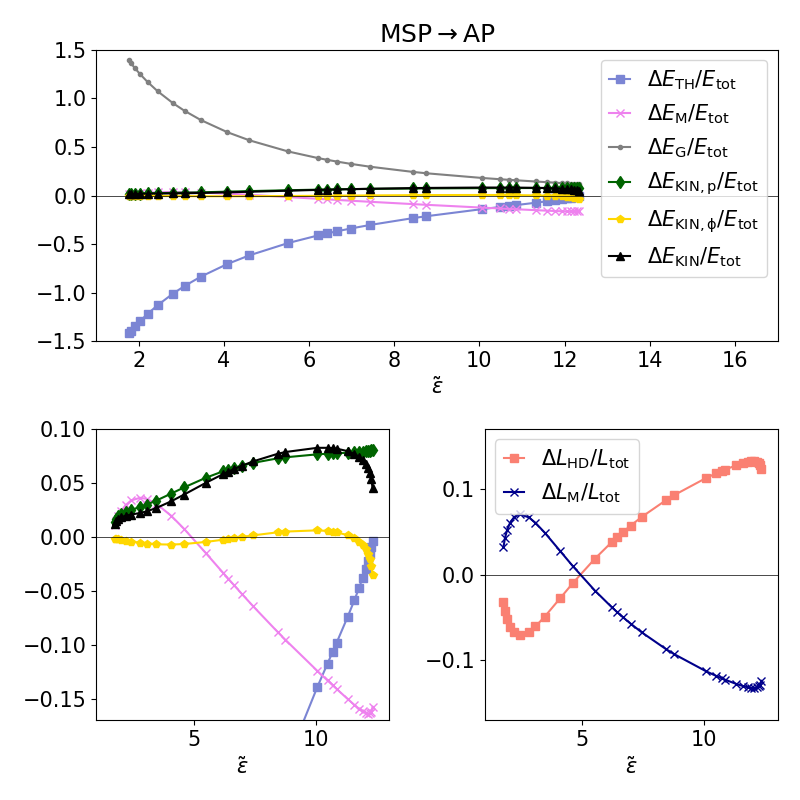}\hfill
\includegraphics[width=1\linewidth]{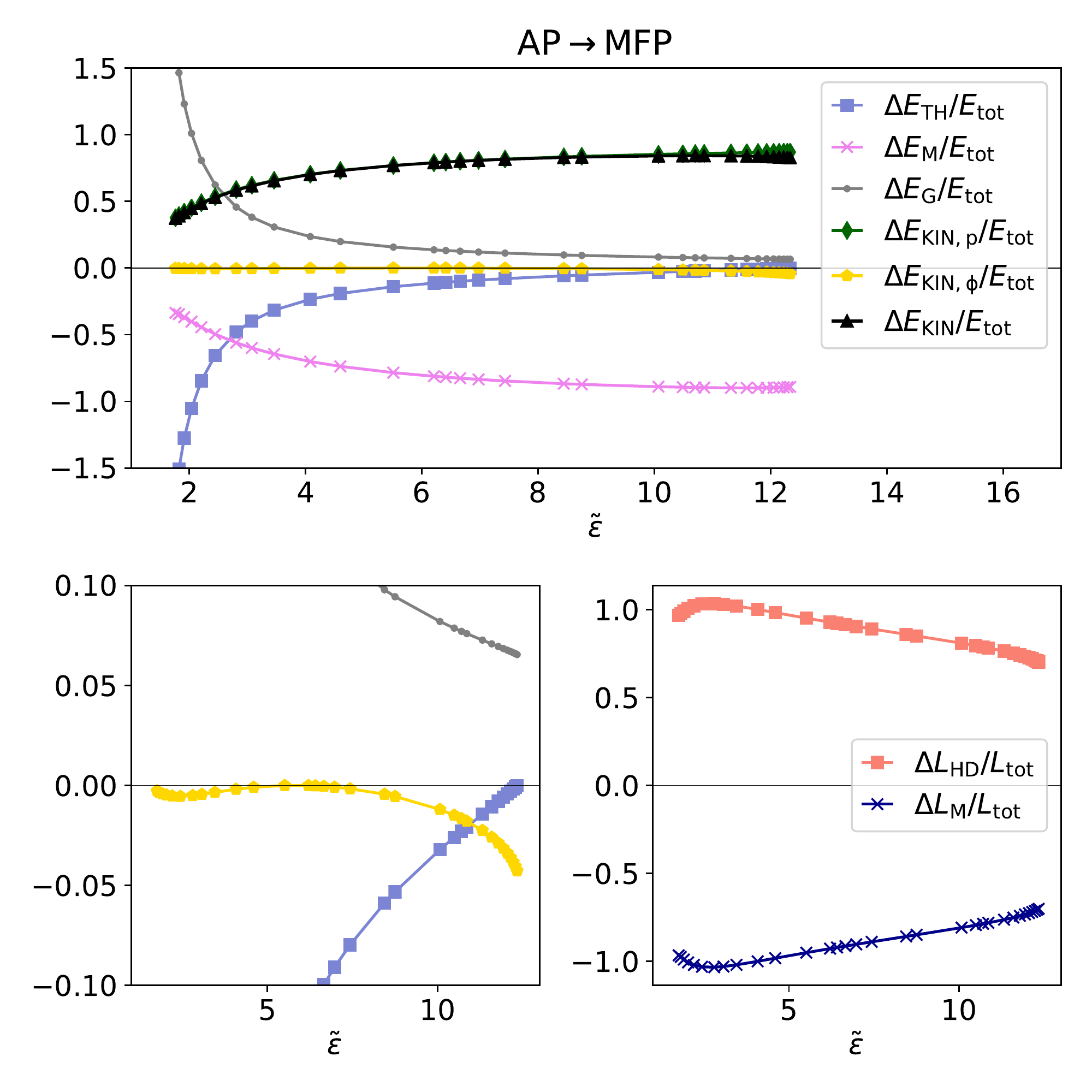}
\caption{\emph{Upper panels:} Relative increment of the energy fluxes with respect to the total energy flux, $(E_{\cdot,\rm AP} - E_{\cdot,\rm MSP})/E_{\rm tot}$, versus the total energy rescaled from the MSP to the AP. The small left panel is a zoom around zero. The small right panel shows the relative increment of the components of the angular momentum with respect to the total angular momentum between the MSP and the AP versus the total energy rescaled. The solutions are obtained for $\theta_{\rm A}=65\degree$ and $k_{\rm VTST}=4.0$. \emph{Lower panels:} Same as the upper panel but between the AP and the MFP.}
\label{fig:deltaE}
\end{figure}

Similarly, the acceleration of the flow is also traced by the increase in the poloidal kinetic energy. In Fig.~\ref{fig:deltaE}, we show the relative increment/decrement of the energy fluxes between the MSP and the AP (first phase of the acceleration, top three panels) and the AP and the MFP (second phase of acceleration, bottom three panels) in a transition from hot to cold solutions (low-to-high energy flux). In the first phase of the acceleration, hot solutions are driven by the thermal energy which suffers the largest decrement. However, as highlighted by the zoom around zero, a fraction of the thermal energy is transferred to the magnetic energy, which is increasing for hot solutions with energy fluxes $< 5$. This behaviour is followed closely by the relative increment/decrement of the components of the angular momentum. For these hot solutions the hydrodynamical component of $L$ decreases, while the magnetic component increases, showing that the angular momentum of the gas is transferred to the angular momentum of the magnetic field.  Such additional channel of energy transfer has been seen in simulations such as e.g. \citet{Komissarov:2009,Cayatte:2014} and in Paper I. This effect is seen as well in the bottom panel of Fig.~\ref{fig:energyfluxes} as a small rise in the magnetic energy around the AP. As the jet models move to higher-energy configurations, the magnetic energy increases while the thermal energy is still important, leading to an increasing poloidal kinetic energy. The peak of the poloidal kinetic energy occurs in correspondence to $\Delta E_{\rm M}/E_{\rm tot} \simeq \Delta E_{\rm TH}/E_{\rm tot}$. Then, it decreases again due to a decrease in toroidal kinetic energy. In the second phase of the acceleration, the thermal energy still dominates for hot low-energy solutions. Equipartition/MC and cold solutions instead are accelerated all the way from the MSP to the MFP by the magnetic field. In the upper part of the jet, the relative increment of the components of the angular momentum do not change sign and the magnetic angular momentum is always transferred to the gas component.

\begin{figure}
\centering
\includegraphics[width=1\linewidth]{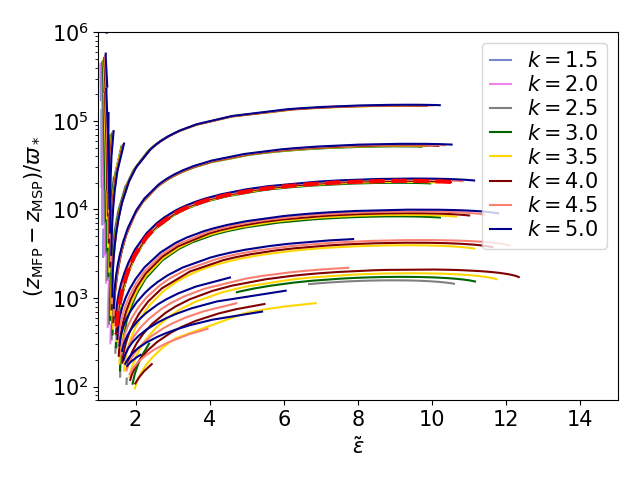}\hfill
\includegraphics[width=1\linewidth]{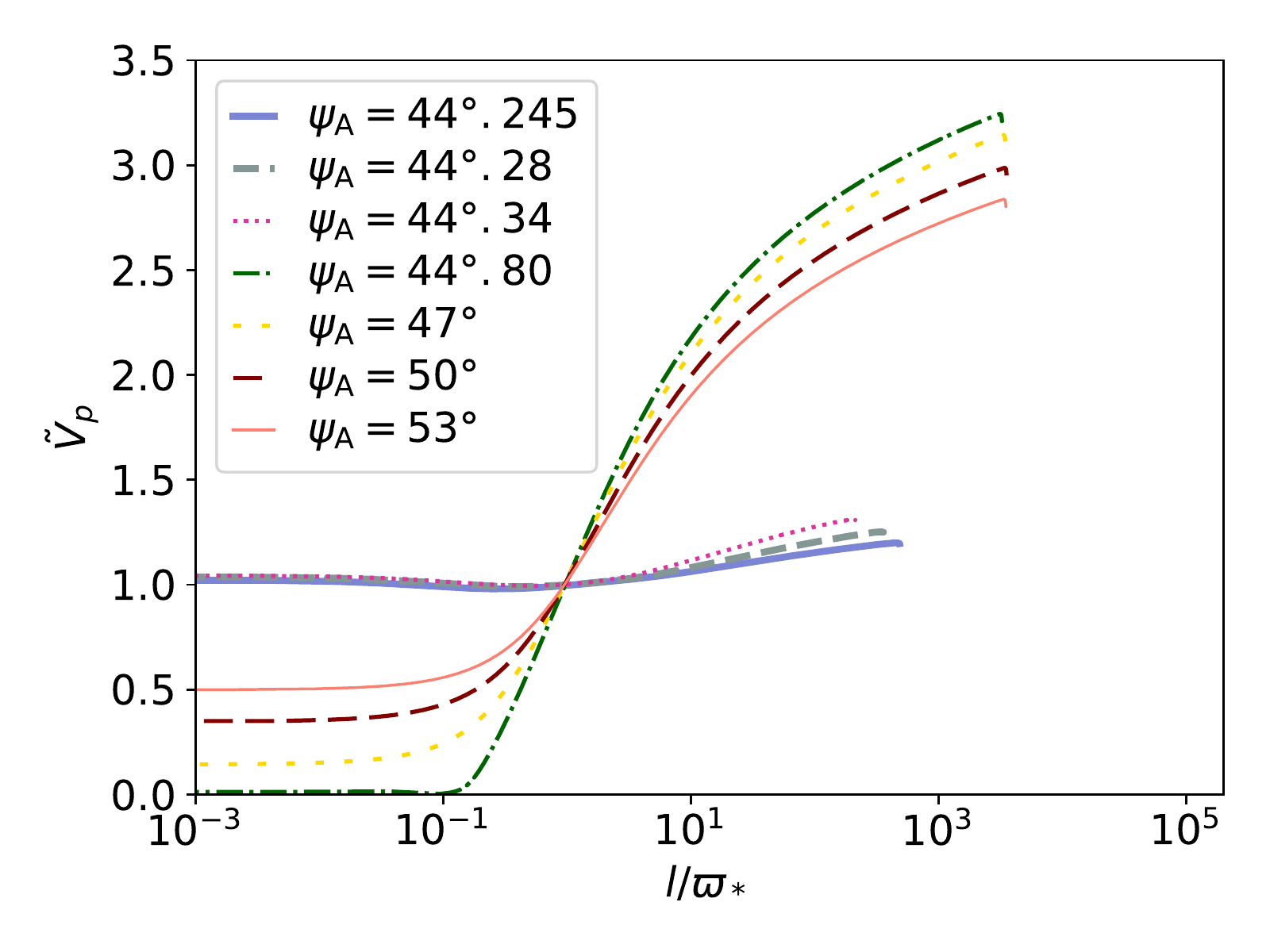}
\includegraphics[width=1\linewidth]{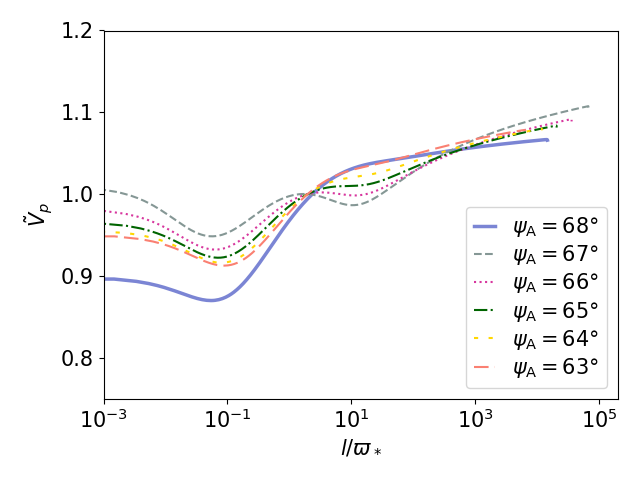}
\caption{\emph{Upper panel}: Distance between the MFP and MSP vs total energy rescaled $\tilde\epsilon = 2 E_{\rm tot}/V_{p,\rm A}^2$. The lines are connecting solutions with constant angular position of the AP ($\theta_{\rm A}$) and the colours mark different values of the mass loss parameter $k_{\rm VTST}$. The red dashed line is a series of solutions obtained with $\theta_{\rm A} = 50\degree$ and $k_{\rm VTST}=4.0$. \emph{Middle panel}: Poloidal velocity scaled with the Alfv\'en poloidal velocity ($\tilde V_p = V_p/V_{p,\rm A}$) along the streamline for a set of solutions with $\theta_{\rm A} = 50\degree$ and $k_{\rm VTST}=4.0$. The solutions in this plot have a total energy flux rescaled ($\tilde \epsilon$) going from 1.5 to 10.5 and distances between the MSP and MFP changing by a factor of $\sim50$. \emph{Bottom panel}: Same plot as the middle one, but for $\theta_{\rm A} = 30\degree$. The solutions in this plot have roughly a constant total energy flux rescaled ($\tilde \epsilon\sim1.3$) and distances between the MSP and MFP changing by one order of magnitude, going from $7900$ to $72500$.The second solutions in the middle and bottom panel have the largest $\Delta z$, but not the largest $\tilde\epsilon$.}
\label{fig:deltaZ}
\end{figure}

Moreover, since the downstream portion of the MFP might be already affected by a shock given by the loss of causal contact with the flow upstream, we take as a proxy the total jet length the distance between the MSP and the MFP. The top panel of Fig.~\ref{fig:deltaZ} shows that low energy solutions can be as short as $10^2/\varpi_*$ and as long as $10^6/\varpi_*$. As the total energy increases this interval narrows by $\sim$2 order of magnitude ($10^3-7\times10^4$). We note that when the streamlines become more vertical (increasing $\psi_{\rm A}$), this leads to a decrease in total energy (the lines in the plot are drawn for constant angular position of the AP, $\theta_{\rm A}$), while increasing $\theta_{\rm A}$  (from top to bottom) makes the $\Delta z$ decrease. If we were to focus on one of the most extended lines across the energy range, for instance the red dashed line that is for an intermediate constant value of the angular position of the AP ($\theta_{\rm A} = 50\degree$) and a fixed mass loss parameter ($k_{\rm VTST} = 4.0$), we would see a correlation between the distance between the MSP and the MFP and the total energy: the higher the energy the larger is the distance, until it reaches an almost constant length ($\sim 20000/\varpi_*$). We note that the maximum of $\Delta z$ does not coincide though with the highest energy in the line. Therefore, beyond a certain total energy, the jets do not grow taller, but their $\Delta V/V_{p,MSP}$ increases as shown in the middle panel of Fig.~\ref{fig:deltaZ} and in  Fig.~\ref{fig:deltaV}. Low-energy hot solutions increase in length by a factor of 10 as the collimation angle,$\psi_{\rm A}$, increases, maintaining their velocity increment roughly constant (bottom panel of Fig.~\ref{fig:deltaZ}).

%

\begin{figure}
\centering
\includegraphics[width=1\linewidth]{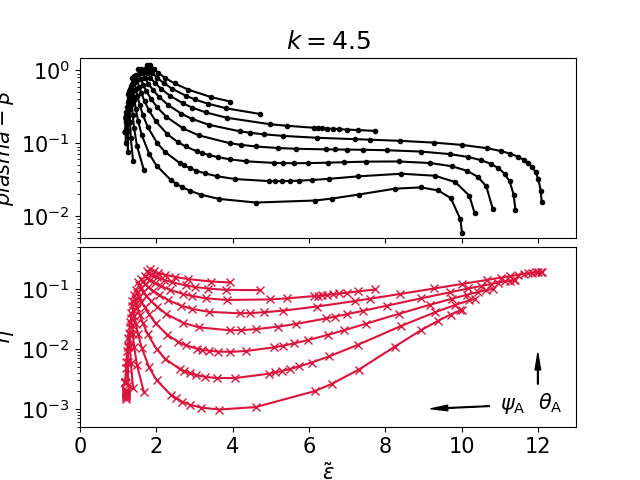}\hfill
\caption{Plasma-$\beta$  (black lines with dots, top panel) and mass load $\eta$ (red lines with crosses, bottom panel) vs total energy flux rescaled for all the solutions with $k_{\rm VTST} = 4.5$ at the MSP. Each line connects solutions with constant $\theta_{\rm A}$ and increasing $\psi_{\rm A}$. The arrows show the approximate direction of increasing $\theta_{\rm A}$ and $\psi_{\rm A}$.}
\label{fig:plasmaBmassloadKconst}
\end{figure}

\begin{figure}
\centering
\includegraphics[width=1\linewidth]{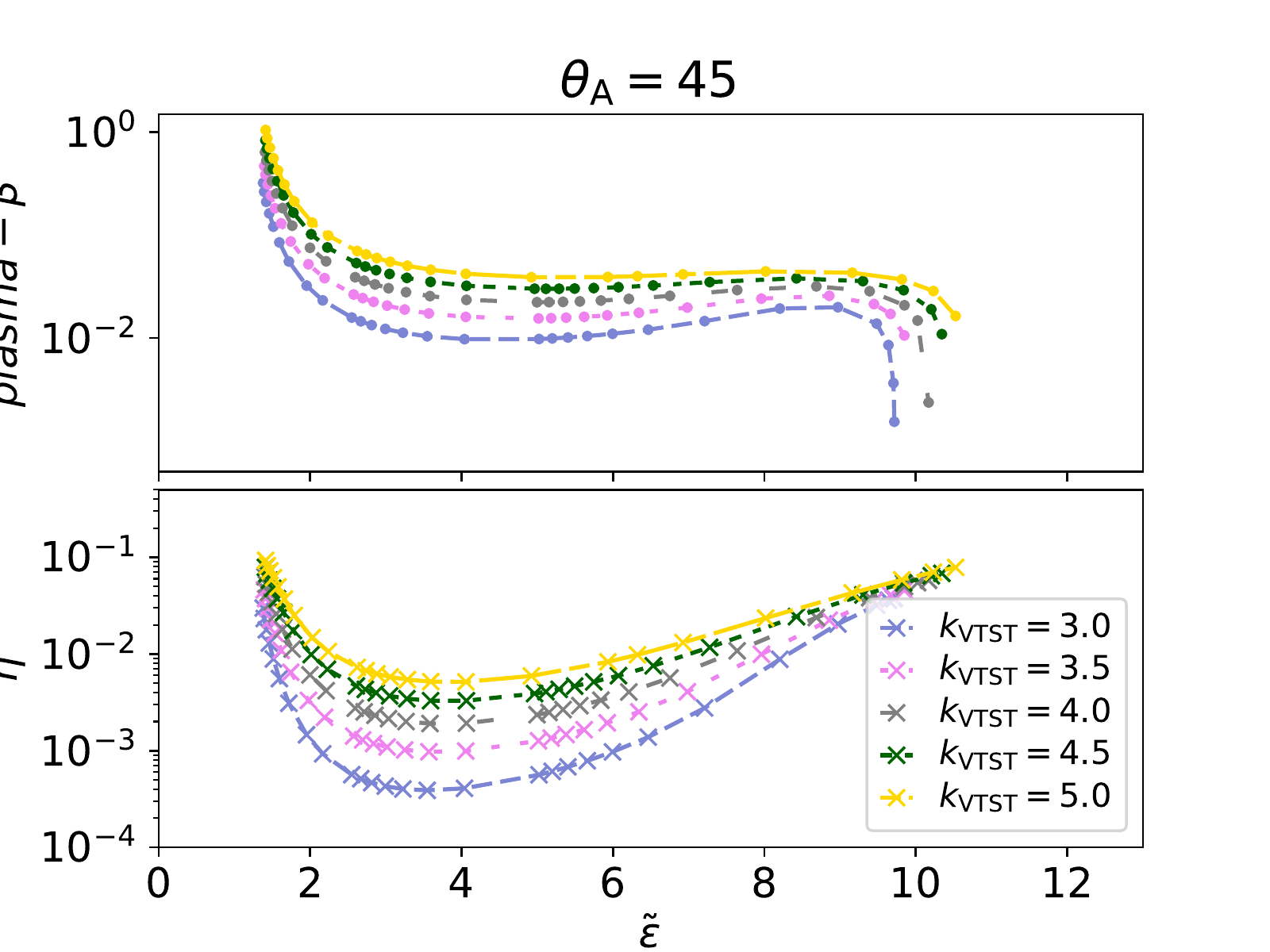}\hfill
\caption{Plasma-$\beta$  (lines with dots, top panel) and mass load $\eta$ (lines with crosses, bottom panel) vs total energy flux rescaled for all the solutions with $\theta_{\rm A}=45\degree$ and $k_{\rm VTST} =3.0\, {\rm( blue)},3.5\, {\rm(pink)},4.0\, {\rm( gray)},4.5 \, {\rm( green)}, 5.0 \, {\rm( yellow)}$ at the MSP. Each line connects solutions with constant $\theta_{\rm A}=45\degree$ and varying $\psi_{\rm A}$.}
\label{fig:plasmaBmassloadTconst}
\end{figure}

Lastly, we discuss the variation of the plasma-$\beta$ and mass load $\eta$ at the MSP, which we identify with the jet base as discussed above. 
These two quantities are given by
\begin{equation}
\beta = \frac{P}{B^2/8\pi} \qquad {\rm and} \qquad \eta =  \frac{4\pi \rho V_p\varpi \Omega}{B_p^2}, 
\end{equation}
following the definitions of e.g. \citet{Anderson:2005,Spruit:1996}.
Since the general trend is the same within subsets of solutions with constant $k_{\rm VTST}$, we present here the series of solutions obtained for $k_{\rm VTST}=4.5$ and for Alfv\'en position angle, $\theta_{\rm A}$, going from $10\degree$ to $75\degree$ roughly from the bottom up (Fig.~\ref{fig:plasmaBmassloadKconst}) and then discuss how they change for increasing $k_{\rm VTST}$ at constant $\theta_{\rm A}$(Fig.~\ref{fig:plasmaBmassloadTconst}).
Solutions in both figures have increasing collimation angle, $\psi_{\rm A}$, along each line from left to right (from $28\degree$ to $83\degree$ in Fig.~\ref{fig:plasmaBmassloadKconst} and from $48\degree$ to $57\degree$ in Fig.~\ref{fig:plasmaBmassloadTconst}).
In Fig.~\ref{fig:plasmaBmassloadKconst}, we see that thermally-dominated, low-energy-flux solutions have the largest plasma-$\beta$ ($\sim1$). Then as the collimation angle, $\psi_{\rm A}$, decreases, the energy flux increases and the plasma-$\beta$ experiences a first decrease. For the Alfv\'en angular positions for which more $\psi_{\rm A}$ values are allowed, we see the plasma-$\beta$ remaining constant for many consecutive solutions of increasing energy flux. However, when the solutions become magnetically-dominated, the plasma-$\beta$ has a drop.
The mass load has a minimum which coincides with the beginning of the plateaux of the plasma-$\beta$, to again rise to higher energy fluxes.
The relatively large mass load of the low energy flux solutions is due to high-density of the gas, while a similar  value is reached for the high energy flux solutions because the magnetic field is more tightly wound up ($|B_\phi/B_p| >1$, see e.g.  \citealt{Anderson:2005,Spruit:1996}).

In Fig.~\ref{fig:plasmaBmassloadTconst}, we show how the same quantities vary in relation to an increase in $k_{\rm VTST}$. 
The plasma-$\beta$ and the mass load, $\eta$, show a similar behaviour with respect to the mass loss parameter, $k_{\rm VTST}$: the larger is $k_{\rm VTST}$ the larger is the plasma-$\beta$ and the mass load. However, we notice that $\eta$ has a weaker dependence on $k_{\rm VTST}$ both at low and high energy fluxes, while the plasma-$\beta$ responds to a change in $k_{\rm VTST}$ more homogeneously across the energy flux interval.

\begin{table}
\centering
\setlength\tabcolsep{4.5 pt}
\setlength\extrarowheight{5pt}
 \caption{Parameters of the solutions used in Sec.~\ref{ssec:hotcold}. The solutions have the following common parameters $k_{\rm VTST}=3.0$,  $\theta_{\rm A}=60\degree$, $\Gamma=5/3$ and $F=0.75$. The Cold Jet and the Hot Jet models are the extremes of the series, while the MC Jet is an intermediate one which is closest to the classical magneto-centrifugal jets encountered in the literature.}
 \begin{tabular}{l | c c c c c}
 \hline
 Model & $\mu_{\rm VTST}$  & $\lambda_{\rm VTST}$ &  $\theta_{\rm MFP}$ &  $\theta_{\rm MSP}$ & $\psi_{\rm A}$ \\
  \hline
  Cold Jet & $7.58\times 10^{-2}$  & 17.3853    &   0.11803   &    1.2462   &  37.07  \\
    \hline
  MC Jet & 0.5396 &  16.8723 &   0.11778 &   1.2578 &   37.09   \\    
    \hline
  Hot Jet & 6.5510 &  1.6861 &    0.12347 &   1.3852 &   46.00 \\
    \hline
  \end{tabular}
  \label{tab:extremes}
  \end{table}

\subsection{Hot and cold jets}
\label{ssec:hotcold}

\begin{figure}
\centering
\includegraphics[width=0.9\linewidth]{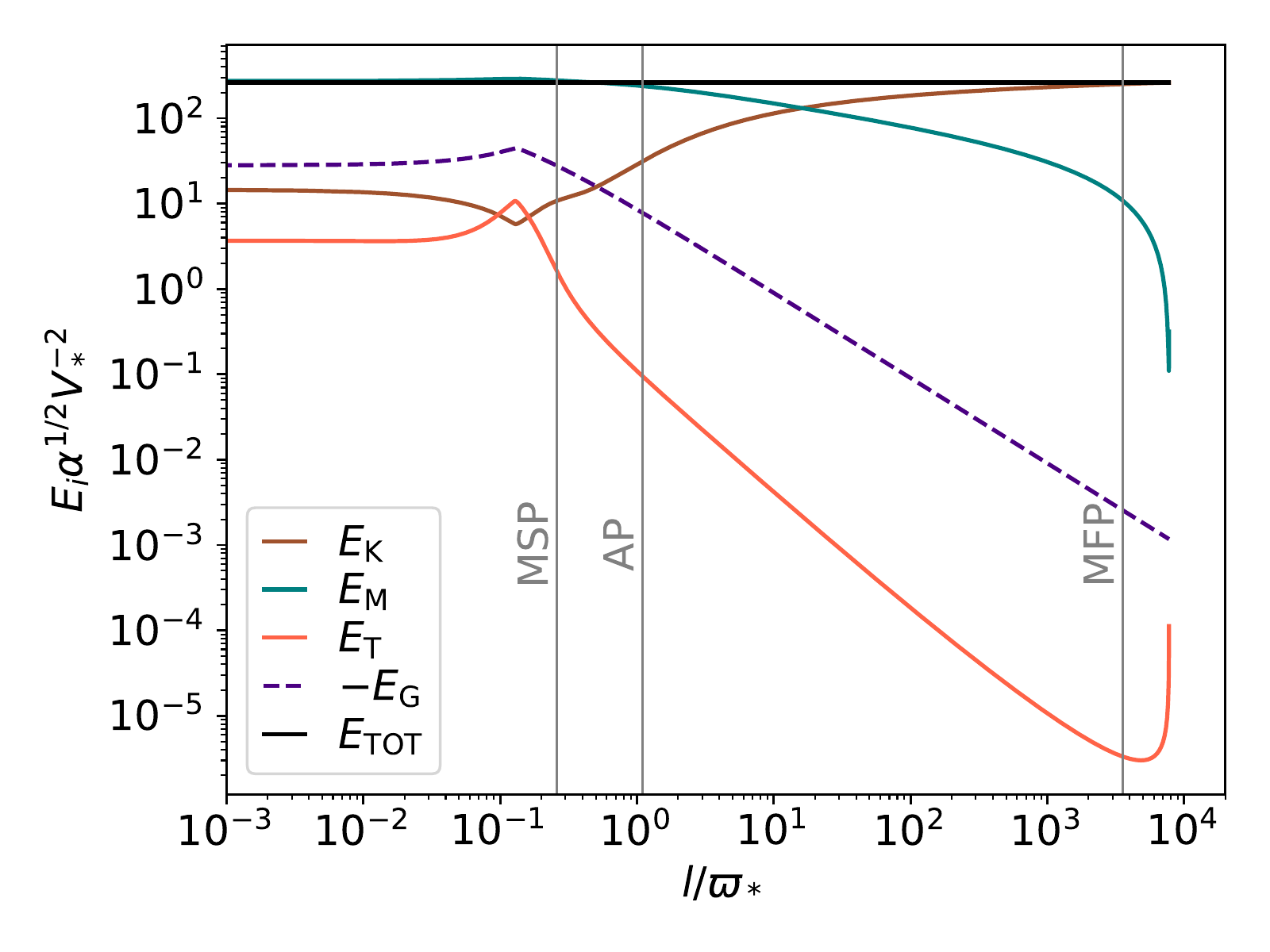}\\
\includegraphics[width=0.9\linewidth]{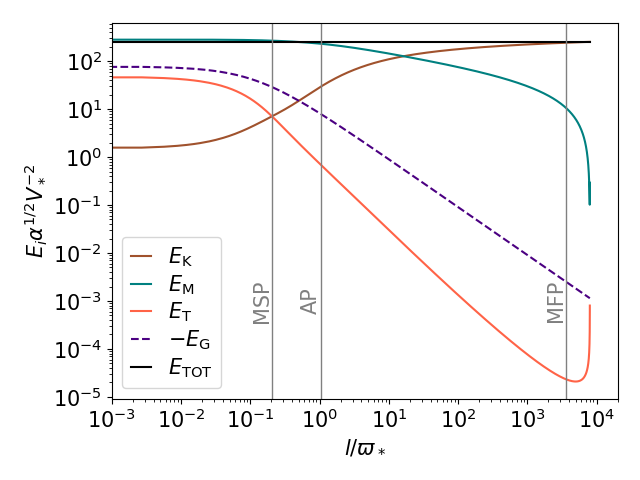}\\
\includegraphics[width=0.9\linewidth]{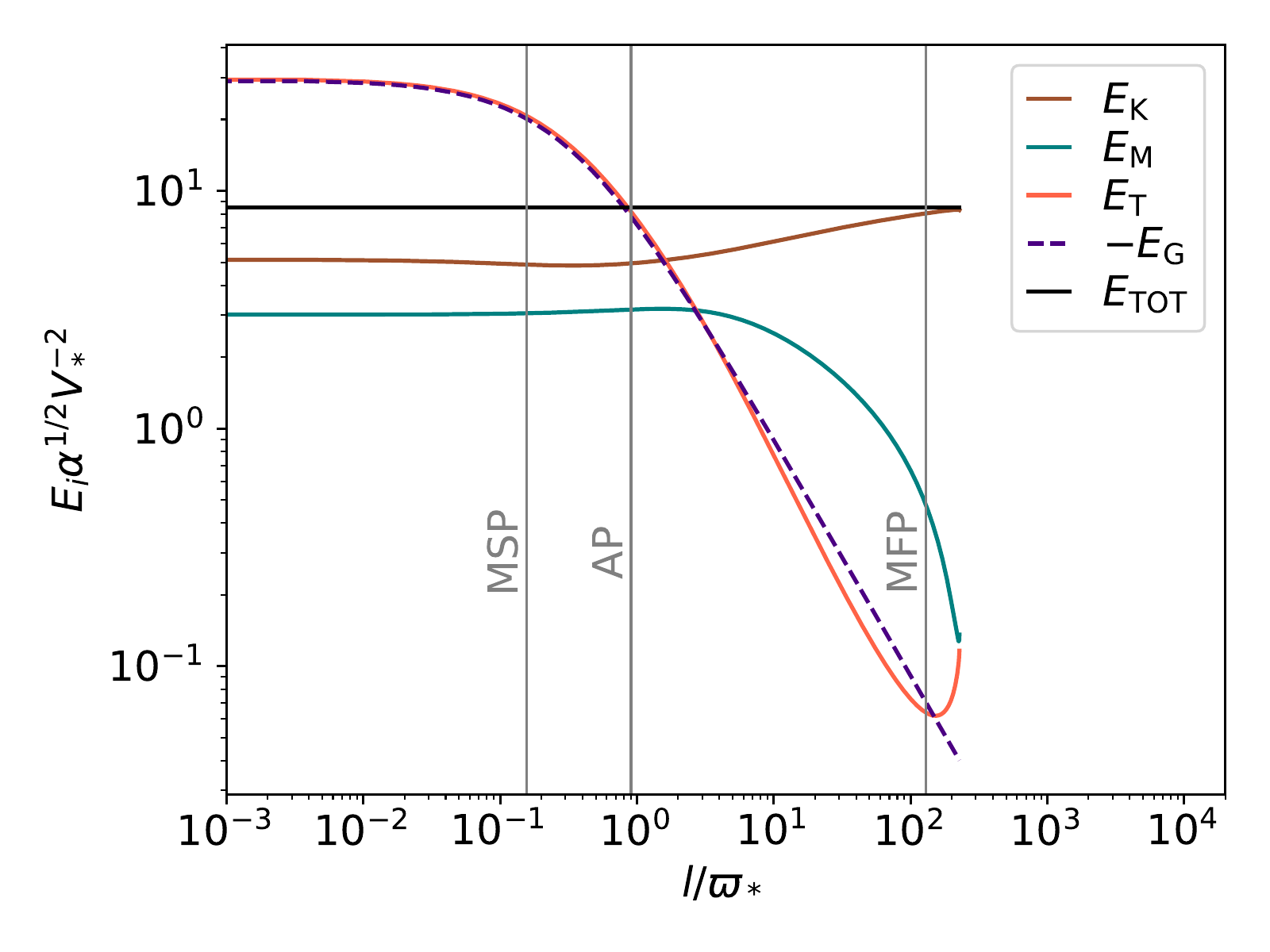}
\caption{Energy fluxes along the streamline. The green line is Poynting flux, the pink line is the enthalpy energy flux, the brown line is the kinetic energy flux and the dashed purple line is the gravitational energy with reversed sign. The total energy flux is shown as a solid black line. Note that the $y$-axis scale is different in the three plots, while the x-axis scale is the same. From top to bottom the solutions go from cold to hot. The parameters are listed in Tab.~\ref{tab:extremes}.}
\label{fig:energyfluxes}
\end{figure}

To illustrate the qualitative changes of the outflow properties along a series of solutions for increasing collimation angle $\psi_{\rm A}$, we describe the transition looking at the two extreme solutions plots of the components of the Bernoulli equation (Eq.~\ref{eqn:BernConst}) and an intermediate one which resembles a more classical magneto-centrifugally launched jet. We will refer to these solutions as Cold, magneto-centrifugal (MC) and Hot Jet models and list their parameters in Tab.~\ref{tab:extremes}.
As shown in Fig.~\ref{fig:energyfluxes}, the energy fluxes along the poloidal direction are substantially different going from the Cold (\emph{upper panel}) to the Hot (\emph{lower panel}) Jet solution.
The cold jet has a high Poynting-to-enthalpy flux ratio. The magnetic energy is then converted into kinetic energy downstream of the AP. Upstream of the MSP, all the energy fluxes are oscillating, following the oscillations of the radial profile of the streamline (See Fig.~\ref{fig:streamlines}). The intermediate MC jet solution has qualitatively the same characteristics of the cold one, but the oscillations are gone.
The hot jet has an uneventful behaviour of the energy fluxes along the streamline. The enthalpy is dominant and roughly equal to gravity in absolute value and opposite in sign. Right after the AP, initially the thermal energy flux is the main source of energy being transformed into kinetic energy and into magnetic energy, which shows a small increase, as discussed in Sec.~\ref{ssec:trends}. Then, the magnetic energy flux takes over the final acceleration. For constant mass loss parameter, $k_{\rm VTST}$, the total energy flux is $\sim 2$ orders of magnitude larger for the cold jet. This larger energy reservoir allows the cold jet to extend in length a factor of $\sim$100 more than the hot jet, when the same reference scale length, $\varpi_*$ is applied.

\begin{figure}
\centering
\includegraphics[width=0.9\linewidth]{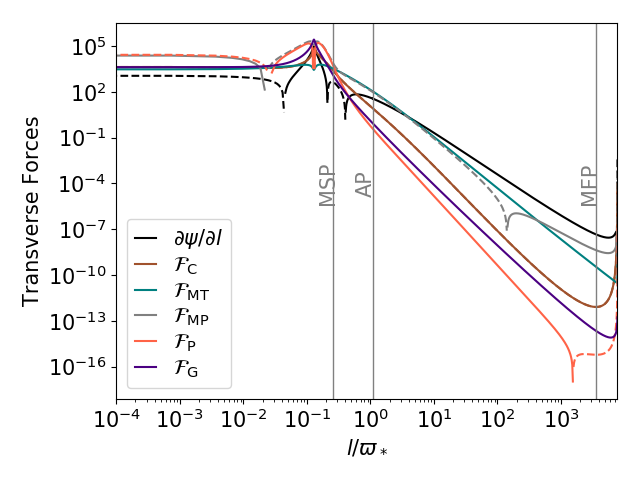}\\
\includegraphics[width=0.9\linewidth]{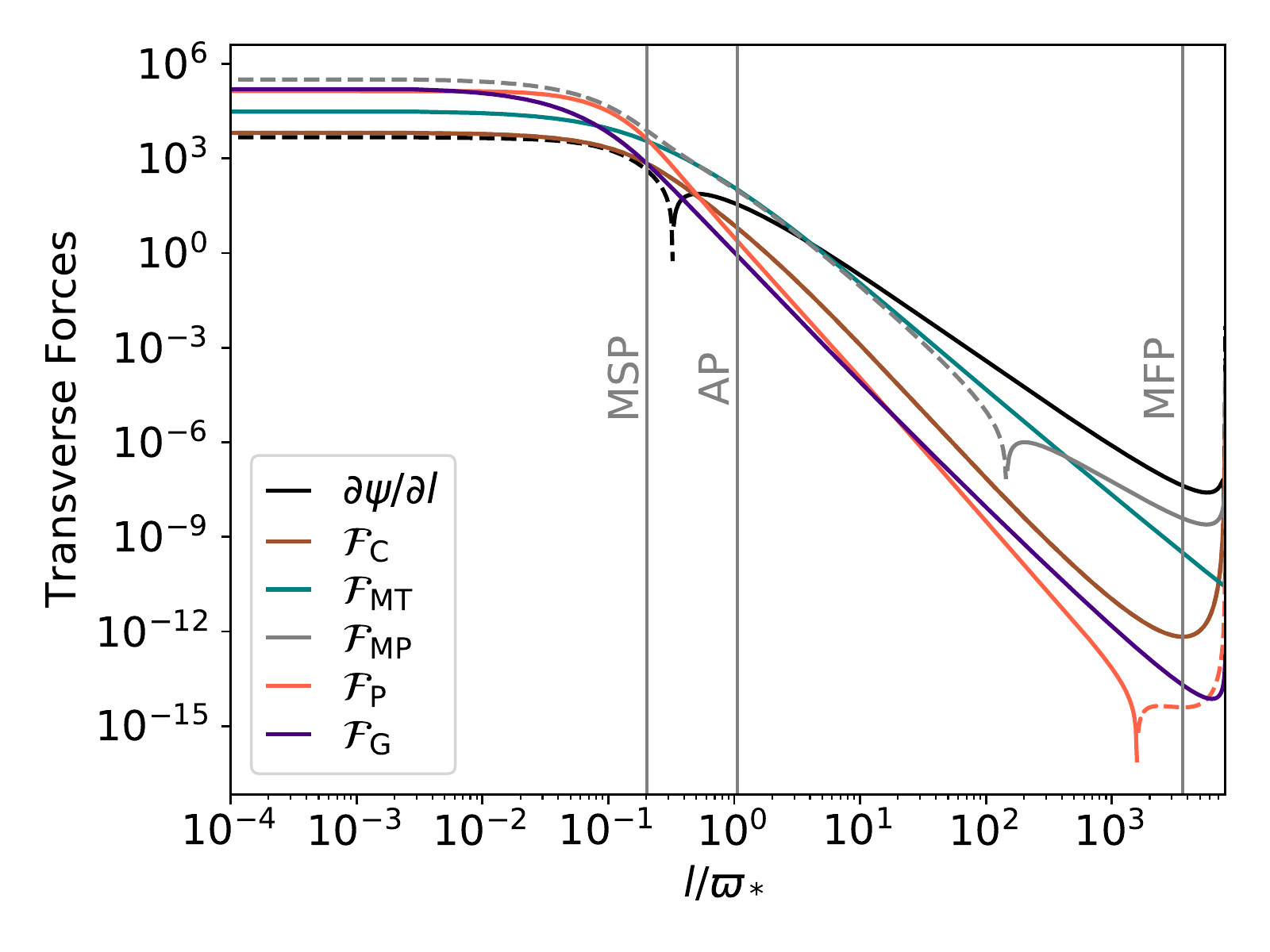}\\
\includegraphics[width=0.9\linewidth]{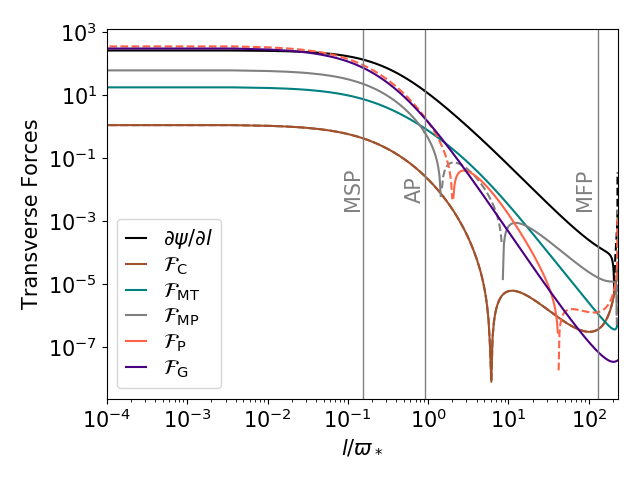}
\caption{Forces perpendicular to the streamline. The black line shows the derivative of the collimation angle along the streamline. The lines are solid when the force is providing collimation (positive) and dashed when it is instead decollimating (negative). From top to bottom the solutions go from cold to hot. The parameters are listed in Tab.~\ref{tab:extremes}}
\label{fig:transfieldF}
\end{figure}

\begin{figure}
\centering
\includegraphics[width=0.9\linewidth]{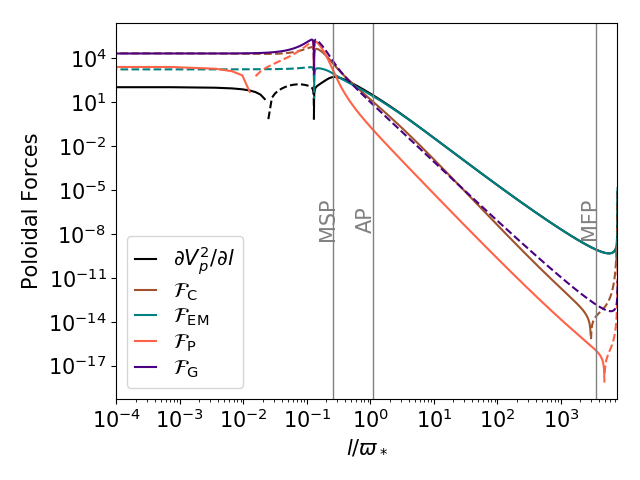}\\
\includegraphics[width=0.9\linewidth]{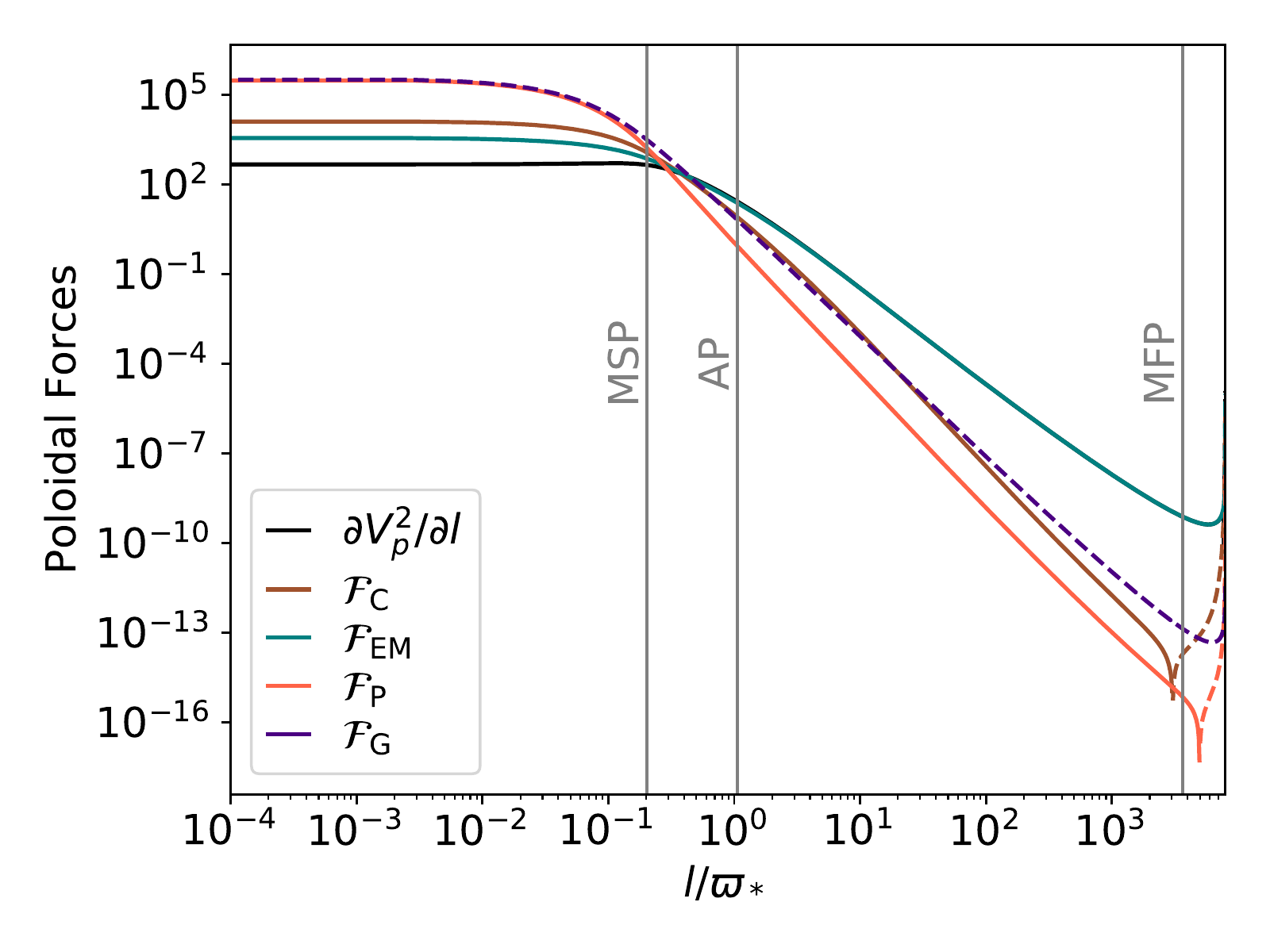}\\
\includegraphics[width=0.9\linewidth]{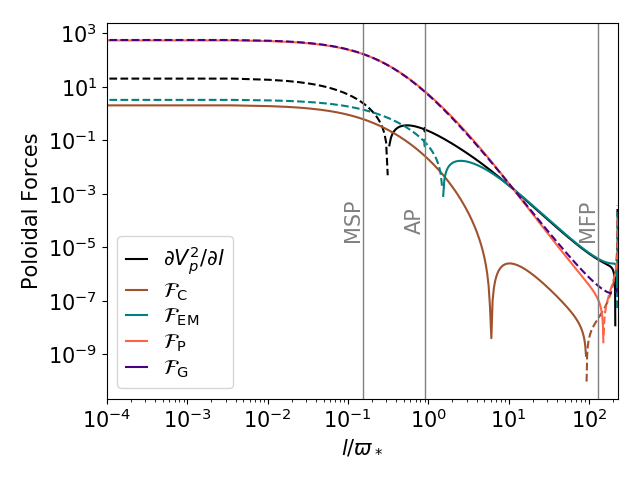}
\caption{Forces along the streamline. The black line is the poloidal acceleration along the streamline. The lines are solid when the force is providing collimation (positive) and dashed when it is instead decollimating (negative). From top to bottom the solutions go from cold to magneto-centrifugal to hot. The parameters are listed in Tab.~\ref{tab:extremes}.}
\label{fig:poloidalF}
\end{figure}

The forces acting along ($\hat b$) and perpendicular  ($\hat n$) to the streamline highlight the transition from cold to hot jet configurations. Here we give the compact form of the forces in both direction, while we provide the full derivation in Appendix~\ref{app:forces}.
\begin{align}
\hat b:&  \frac{\rho}{2} \frac{\partial V_p^2}{\partial l}   =  \rho V_\phi^2 \frac{\cos(\psi)}{\varpi} - \frac{\partial P}{\partial l} + \rho \frac{\partial }{\partial l}\left(\frac{\mathcal{GM}}{r}\right) \nonumber\\ 
& \qquad \qquad - \frac{1}{8\pi}\frac{\partial B_\phi^2}{\partial l} - B_\phi^2\frac{\cos(\psi)}{4\pi\varpi} \label{eqn:poloidalForces}\\
\hat n: &  \left(\rho V_p^2 -  \frac{B_p^2}{4\pi}\right)\frac{\partial \psi}{\partial l}  =  + \rho V_\phi^2 \frac{\sin(\psi)}{\varpi} - \frac{\partial P}{\partial n} + \rho \frac{\partial }{\partial n}\left(\frac{\mathcal{GM}}{r}\right) \nonumber\\ 
& \qquad \qquad  \qquad \qquad - \frac{1}{8\pi}\frac{\partial }{\partial n}\left(B_p^2 + B_\phi^2\right)  +  B_\phi^2\frac{\sin(\psi)}{4\pi\varpi} \label{eqn:transvForces}
\end{align}
The term on the lhs of the Eq.~\ref{eqn:poloidalForces} is the acceleration along the streamline, the first term on the rhs is the centrifugal force, the second term is the gas pressure force, the third term is the gravitational force and the last two terms are the magnetic pressure gradient and the magnetic tension.
On the lhs of Eq.~\ref{eqn:transvForces} there is the derivative of the angle $\psi$ along the streamline. The inverse of this derivative is also called the collimation radius, $R_c = (\partial\psi/\partial l)^{-1}$. 
On the rhs there are: the centrifugal force, the gas pressure force, the gravitational force and the magnetic pressure gradient and the magnetic tension.
In the following discussion, we refer to accelerating/collimating forces when such terms are positive, and to decelerating/decollimating forces when they are negative.
In Fig.~\ref{fig:transfieldF}, we show the forces perpendicular to the streamline and in Fig.~\ref{fig:poloidalF} the forces along the streamline for the same three solutions. 

\begin{figure}
\centering
\includegraphics[width=0.9\linewidth]{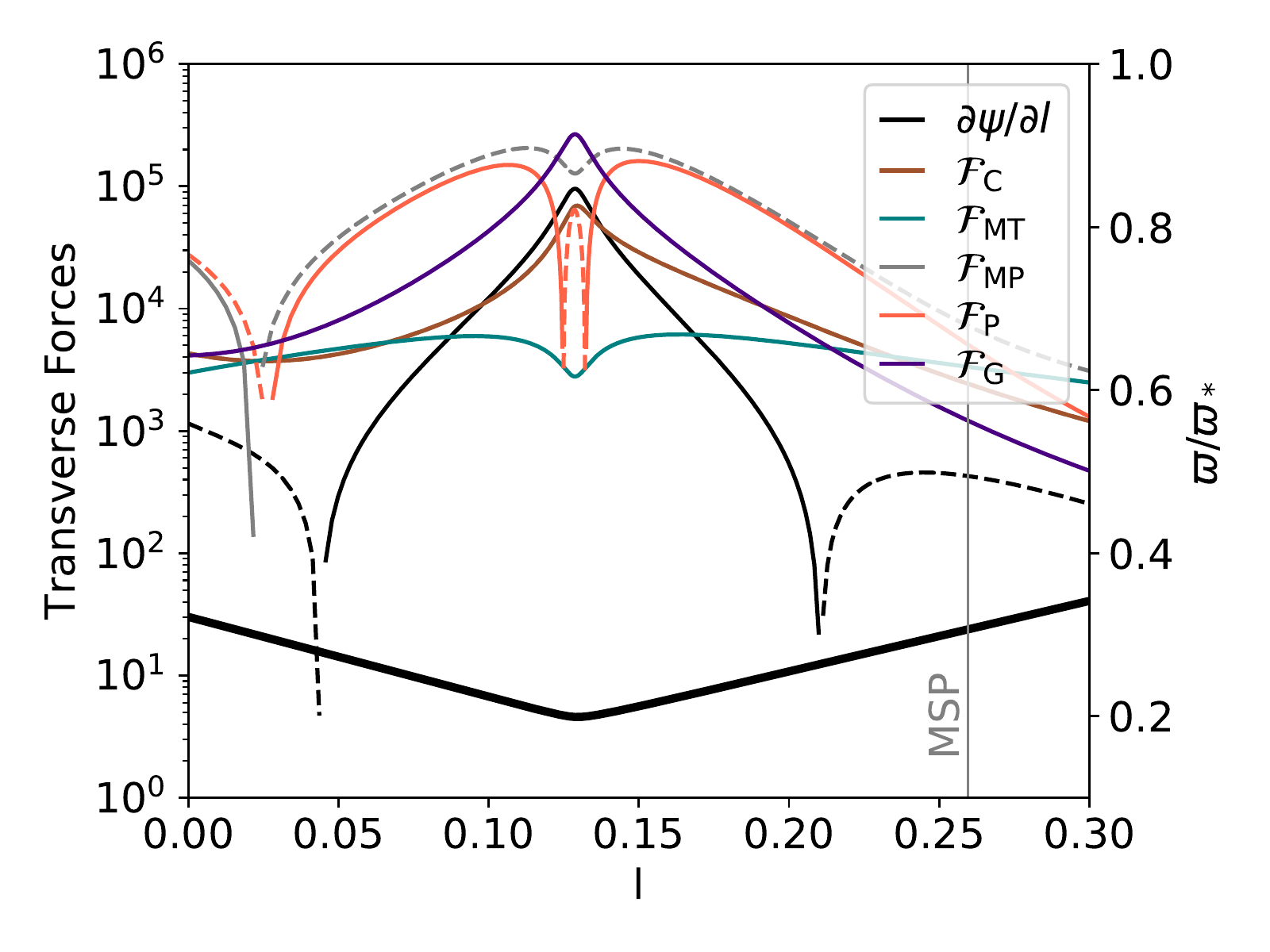}\\
\includegraphics[width=0.9\linewidth]{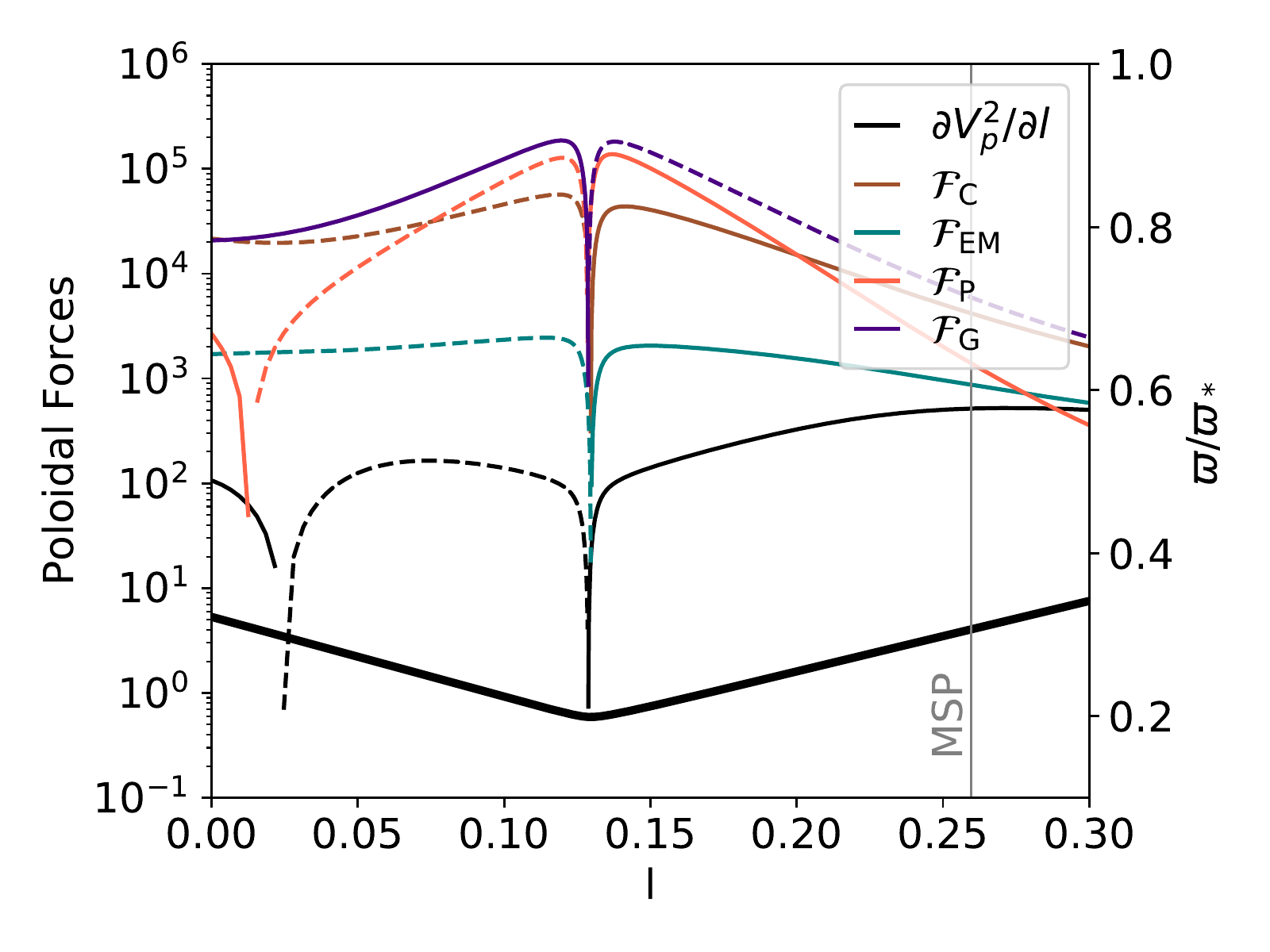}
\caption{Zoom of the forces along the streamline (bottom) and perpendicular (top) to it for the Cold Jet model (Tab.~\ref{tab:extremes}). The thick solid black line is the streamline.}
\label{fig:zoomF}
\end{figure}

The cold jet has a troublesome start, since it lacks a vertical velocity component that allows for a straightforward launching (top panel in Fig.~\ref{fig:zoomF}). 
At the very beginning, the jet is decollimating ($\partial \psi/\partial l$ < 0, black thin line) under the action of the gas pressure force (pink line). Soon, the gas pressure gradient changes sign and together with the other positive forces, i.e. gravity (purple line), centrifugal (brown line) and magnetic tension (teal line), is collimating the jet against magnetic pressure gradients. Around the peak of gravity and the centrifugal force, the pressure gradient becomes negative but smaller in modulus, resulting in a converging streamline (thick solid black line). After that the previous configuration of the forces is mirrored to the right side of the peak, until shortly before the MSP, the streamline starts to decollimate again. However, downstream of the MSP the sign switches again when the magnetic tension becomes dominant, keeping the jet collimated up until also the magnetic pressure gradient becomes positive, about half way between the AP  and the MFP (top panel of Fig.~\ref{fig:transfieldF}).
The MC Jet model shows the same behaviour downstream of the MSP, while it presents no oscillations in the region between the disk and the MSP.
 
The Hot Jet is always collimating. Until the AP, gravity is the main force driving the collimation against the gas pressure gradient that remains negative until past the AP. Beyond this point, the magnetic forces become dominant in keeping the jet focused.

The poloidal forces also presents oscillations in the Cold Jet model, while they do not in the MC and Hot Jet models. 
In the top panel of Fig.~\ref{fig:poloidalF}, we see some more moderate oscillations for the initial segment of the cold jet. In the upstream region of the MSP, the jet is initially slowly accelerating ($\partial V_p^2/\partial l>0$). 
Then the pressure force (pink line) becomes negative and gravity (purple line) is
attracting the fluid back to the centre (the thick solid black line in Fig,~\ref{fig:zoomF} shows the streamline focussing towards the axis), now increasing its speed,
providing acceleration while the poloidal motion of the flow (think black line) is actually decelerating (bottom panel in Fig.~\ref{fig:zoomF}). At the minimum radius of the streamline, all the forces change sign and the jet starts to accelerate driven by a combination of centrifugal (brown) and magnetic force (teal line). Half way between the MSP and the AP, the magnetic force takes over and it will sustain the acceleration for the remaining (and larger) fraction of the jet extent.
In the intermediate jet solution the flow is accelerated by the gas pressure force (and for a small segment just downstream of the MSP by the centrifugal and the magnetic forces) until halfway the MSP and the AP, when the magnetic force drives again the acceleration of the jet until the LRP.
The Hot Jet is instead decelerating until past the MSP, then the pressure provides acceleration working against the gravitational pull. Finally halfway between the AP and the MFP, the magnetic force becomes the dominant accelerating force for the rest of the jet length.




\section{Proof of concept: application to W43A}
\label{sec:app}

In this section we describe how to compare our solutions to an astrophysical source, the water fountain W43A. 

W43A is a pre-planetary nebula (PPN; plural, PPNe), located at a distance of 2.2 kpc from the sun \citep{Tafoya20}, that is thought to be hosting an Asymptotic Giant Branch (AGB) star \citep{ImaiDiamond:2005, Tafoya20}. It has been observed that during the transition from the AGB to planetary nebula (PN) phases, the star’s ejecta change from a roughly spherical symmetric wind to an envelope with a highly non-spherical configuration \citep{BalickFrank:2002}. These non-spherical post-AGB or PPNe envelopes often exhibit (collimated) bipolar outflows and/or jets, which are most likely formed at the time the star leaves the AGB \citep[e.g.][]{ST98}. The origin of the non-spherical outflows around W43A and other PPNe is a matter of debate, and is typically thought to include a common envelope evolution (CEE) phase \citep[e.g.][]{NB06}. It is suggested that W43A also hosts a close companion embedded in the circumstellar envelope of the AGB star, likely a main sequence star or a white dwarf, although such companion has not been directly observed \citep[e.g.][]{Imai02,ImaiDiamond:2005,Tafoya20}. The binary interaction between the two stars is expected to lead to the ejection of the envelope. During this phase, both a circumbinary disk and an accretion disk around the companion can form. It has been proposed that fast outflows, either collimated or wide, can be launched before, during and/or after the common envelope phase, contributing to the evolution of the system by heating and mechanically re-disturbing the material of the envelope, possibly leading to its ejection \citep{Chamandy:2018, Soker:2020}. A scenario in which jets are launched at the onset of the short-lived water fountain phase of W43A life cycle seems plausible considering the current properties of the source \citep{Tafoya20}.

Following the argument that \citet{Sahai:2017} used for the water fountain IRAS 16342-3814, if we were to assume that the radiation pressure is the main force responsible for the launching and acceleration the jets of W43A, we could estimate the timescale for ejecting such radiation-driven jets as
\begin{equation}
\tau_{\rm rad} = \frac{Pc}{L}
\end{equation}
where $P$ is the total momentum, $L$ is the luminosity of the source and $c$ is the speed of light.
The momentum derived from observational constraints is $P\sim 3.06\times 10^{37}$ g cm/s. Adopting a luminosity of 6000 $L_\odot$ given by \citet{DuranRojas:2014}, we obtain a timescale of $\sim 1268$ yr which is almost 20 times larger than the dynamical timescale ($t_{\rm dyn}\sim 65$ yr) estimated by \citet{Tafoya20}. Thus, 
radiation can be ruled out as the mechanism responsible for launching and accelerating the jet.

Several mechanisms have been proposed to produce collimated jets, many of which make use of magnetic fields to drive, or at least to strongly contribute to, the acceleration and collimation of the material from a rotating object, i.e., a star, a compact object or a disk \citep[e.g.][]{Shu:2000,BlandfordZnajek:1977,BlandfordPayne:1982, Ferreira:1997, Parfrey:2016} and a similar contribution has been proposed for PPNe as well \citep[e.g.][]{GarciaSegura:2005}.

\subsection{Observational constraints}
\label{ssec:obsConstr}

\begin{figure}
\centering
\includegraphics[width=0.99\linewidth]{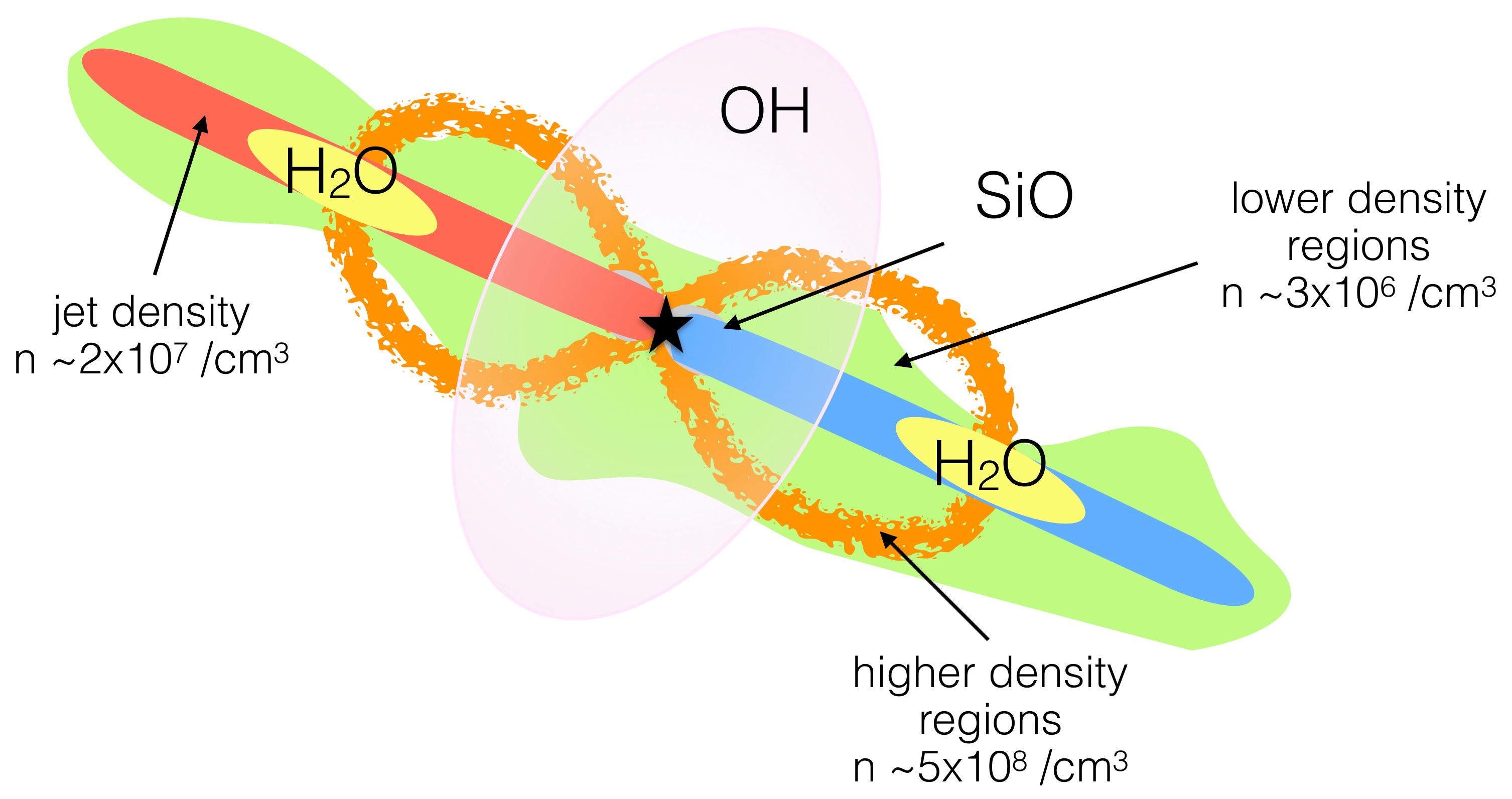}
\caption{Sketch of W43A masers emission regions.}
\label{fig:W43A}
\end{figure}

Recent observations by \citet{Tafoya20} show that W43A possesses a dense ($n\sim2\times10^7$ cm$^{-3}$), collimated ($z/\varpi \sim 20$, where $\varpi$ is the radius of the jet and z is its height) molecular jet. The molecular jet inclination angle with respect to the plane of the sky is 35$\degree$, and its position angle (P.A.; with respect to the north) is 68$\degree$. The jet extends with constant collimation angle out to a distance from the central source of $\approx$1600~AU, and it is surrounded by two lobes of shocked material with a lower density ($n\sim 3\times10^6$ cm$^{-3}$) (see Fig.~\ref{fig:W43A}). 

W43A is known to host maser emission from different chemical species, such as OH, 
H$_{2}$O and SiO. The OH masers are located on an expanding torus of radius $\sim500$ AU with an expansion velocity of $\sim18$ km/s and a velocity separation of $\sim 16$ km/s. The density required for the excitation of the OH masers at that distance is $\sim10^4-10^6$ cm$^{-3}$ \citep{Elitzur:1992}. The H$_2$O maser emission is observed at the two regions where the jet seems to be interacting with the lobes. The H$_2$O maser spots have velocities $\sim 150$ km/s and hydrogen densities  $\sim10^8-10^{10}$ cm$^{-3}$ \citep{Imai02, Vlemmings06a, Vlemmings06b}. SiO masers have also been observed, at $\sim 70$ AU from the star \citet{ImaiDiamond:2005}, and were modelled as an expanding shell of shocked material surrounding a high velocity outflow. The magnetic field in the material surrounding W43A has been measured using observations of the Zeeman splitting of H$_2$O and OH masers \citep{Vlemmings06a,Amiri10}. The magnetic field strength measured in the H$_2$O maser regions is $\sim200$~mG \citep{Vlemmings06a}. 

The magnetic field in the maser regions is likely enhanced, due to compression of the field lines in the shocked interaction region between the jet and the surrounding medium. The H$_2$ number density in the lobes around the jet is estimated to be $3\times10^6$~cm$^{-3}$ and that in the surrounding shell is $5\times10^8$~cm$^{-3}$ \citep[][ Fig.~\ref{fig:W43A}]{Tafoya20}. Using these densities to update the uncompressed magnetic field estimates from \citet{Vlemmings06a} and \citet{Amiri10} and assuming a typical H$_2$O maser region number density of $10^9$~cm$^{-3}$ and a magnetic C-shock, we find a magnetic field strength in the range of $\sim0.6$~mG, when the shock occurs in the lower density material of the lobes, to $\sim100$~mG, if the shock occurs in the denser shell surrounding the lobes. Since the exact maser density is unknown, the uncertainty on these values is large. Although it is unclear exactly which component of the magnetic field is traced by the H$_2$O maser measurements, the linear polarisation direction and evidence of change in sign of the measured magnetic field across the jet indicate that the masers likely probe the toroidal magnetic field \citep{Amiri10}.
We refer to the bipolar high velocity outflow traced by the H$_2$O masers as the \emph{molecular jet} of W43A.

\begin{figure}
\centering
\includegraphics[width=0.99\linewidth]{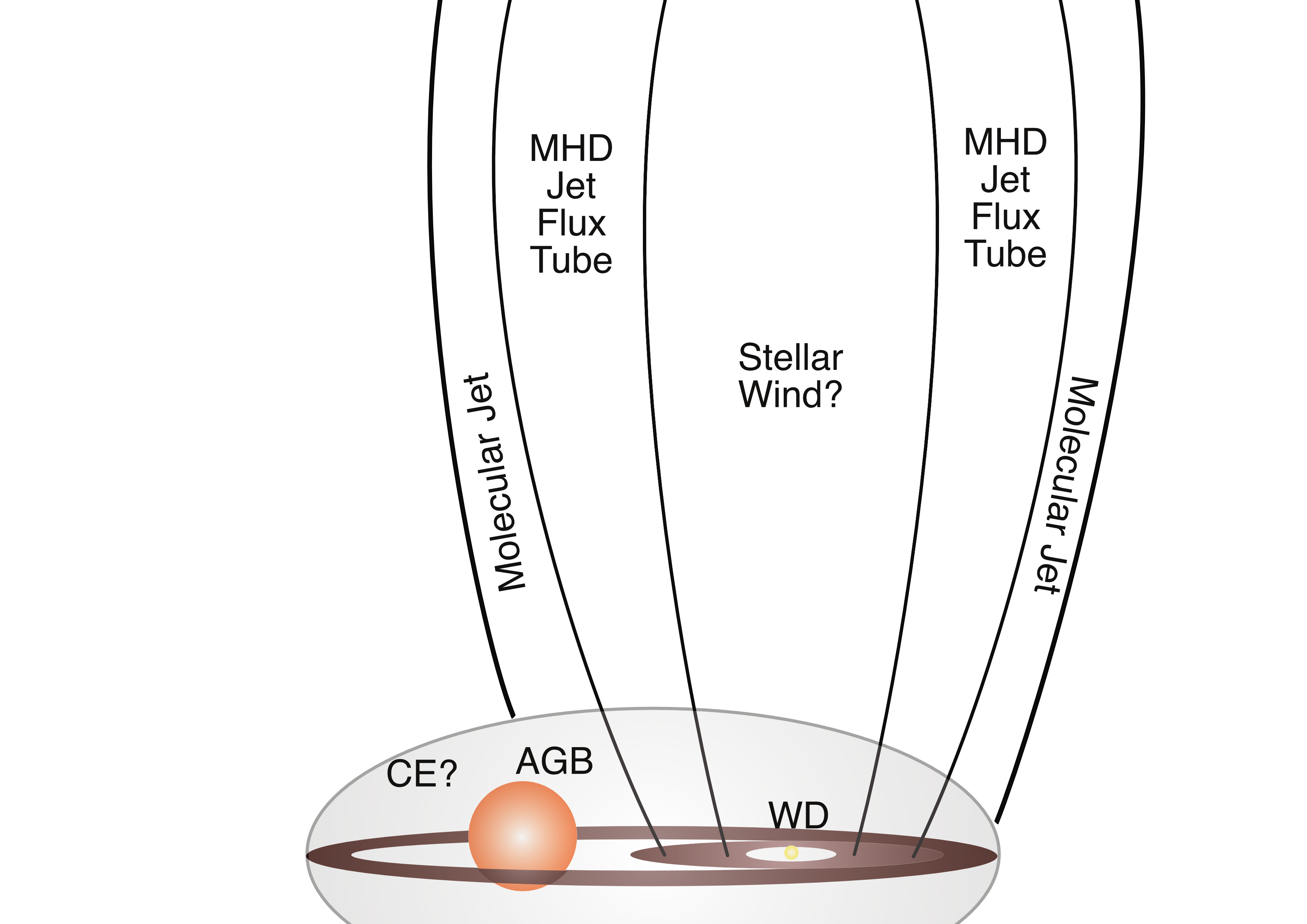}
\caption{Sketch of the central region of W43A.}
\label{fig:W43Asource}
\end{figure}

\subsection{Modelling assumptions}
\label{ssec:modAssumpt}

How such molecular jet is launched and how it maintains its collimation throughout its length has not yet been established. 
We hypothesise that what is shaping the molecular jet of W43A is a \emph{disk-driven MHD jet}. 
More specifically, we assume that a MHD jet is launched by an accretion disk formed around a white dwarf companion ($\mathcal M = 0.6 M_\odot$, $\mathcal R = 0.01 R_\odot$) orbiting around the AGB star (see Fig.~\ref{fig:W43Asource}). 
In this scenario, the MHD jet is accelerated outwards and entrains material from the surroundings, building up a more mass loaded, slower cocoon which is observed as a molecular jet \citep[e.g.][]{Hardee:1996,Rosen:1999}. 
In this paper we show that the properties of disk-driven MHD jets can be very diverse. In order to reduce the allowed range of such properties, we compare our solutions to the observational constraints of the observed molecular jet of W43A. If the MHD jet is driving the molecular jet, its momentum has to be at least equal, or larger, than the momentum carried by the molecules ($P_j \ge P_{j,\rm obs}$). 
Since the way in which the composition of the MHD jet relates to the molecular content is unknown, we assume that its hydrogen number density is at most equal to the hydrogen density estimated from the CO mass ($n \ge n_{\rm obs}$).
With a lower density, the MHD jet is also likely to travel at a faster speed than the water maser spots in that region ($V_{\rm H_{2}O} \ge V_{\rm H_{2}O, obs}$).
Finally, we adopt the full range of the toroidal magnetic field strength ($B_\phi = 0.6-100 $ mG) to look for solutions which have the requirements listed above and extend for 2000 AU.
Given these uncertainties, we choose to compare the momentum rate carried by our solutions with the one estimated with the observational constraints to avoid a direct comparison between densities and velocities where we should instead make more assumptions such as on the ionization fraction or on the intrinsic speed of the jet.
This allow us to predict the general properties of the MHD jet to drive the molecular jet sheath surrounding it.

To estimate the momentum rate from observations, we approximate the molecular jet of W43A as a full cylinder with a length of 2000 AU and a radius of 45 AU. The momentum rate can be estimated as $\dot P = M_j V_{\rm H_{2}O}/t_{\rm dyn}$, which for a jet with a total mass of $\sim 10^{-3}$ M$_{\odot}$, a velocity of 150 km/s and a dynamical timescale of 65 years is $\sim 0.0023$ M$_\odot$/yr km/s. This is equivalent of a total momentum over 65 years of $P\sim 3.06\times 10^{37}$ g cm/s.
We note that this estimated value of the momentum of W43A lies within the range ($10^{35}-10^{40}$ g cm/s) reported by  \citet{BlackmanLucchini:2014} for a sample of pre-planetary nebulae showing high-velocity and extreme high-velocity outflows. 
Finally, we derive the mass loss rate of $\dot M_j = \pi \varpi_0^2\rho V_{\rm H_{2}O} \sim 1.58\times10^{-05}$M$_\odot$/yr.
For these calculations, we have considered a  constant density and velocity along the jet axis and along the jet radius as well. In the next section, we will discuss how to derive averaged quantities in physical units from a single scale-invariant streamline, which is what we call a solution.

\subsection{Scaling of the solutions}
\label{ssec:selection}

We use a sample of roughly 1500 solutions mapping the parameter space and we start by selecting the ones that satisfy the criterion:
\begin{equation}
\left(\frac{\varpi}{z}\right)_{\rm H_{2}O} = \tan(\theta_{\rm H_{2}O}),
\label{eqn:thetaH2O}
\end{equation}
which is equivalent to determining which solutions terminate beyond the H$_2$O spots ($\theta_{\rm LRP} < \theta_{\rm H_{2}O}$).
The quantity that determines $\theta_{\rm H_{2}O}$ is the jet cylindrical radius at $z_{\rm H_{2}O} = 1000$ AU, while we keep the jet height at the water maser spots constant.
Observational constraints on the cylindrical radius at H$_2$O  vary from $\sim10$ AU \citep{Imai02} to $ \sim45$ AU \citep{Tafoya20}. 
We treat the cylindrical radius at H$_2$O as a free parameter within the above interval.

Since our solutions are calculated in dimensionless units, e.g. ($\varpi/\varpi_*, V/V_*, B/B_*, \rho/\rho_*, ...$), the first step to compare them with a physical system is to introduce a characteristic length to scale them. 
We use the cylindrical radius of the jet at the H$_2$O masers spots as reference length to scale the cylindrical radius of each solution, $\varpi$, through the relation 
\begin{equation}
\varpi_* = \left(\frac{\varpi}{G}\right)_{\rm H_{2}O}
\label{eqn:wscale}
\end{equation}
for the reference streamline ($\alpha=1$, see Sec.~\ref{ssec:equations}). 
Once the reference length is fixed, the scaling of the velocity is also defined as
\begin{equation}
V_* =  \sqrt{\frac{\mathcal {G M}}{\kappa_{\rm VTST}^2\varpi_*}}
\end{equation}
for a given central object with mass (for a white dwarf, $\mathcal M = 0.6 M_\odot$).
We use the observational constraints on the toroidal magnetic field component to determine the maximum and minimum $B_*$ as
\begin{equation}
B_* = \frac{B_{\phi,\rm{obs}}}{B_{\phi,\rm{H_{2}O}}},
\end{equation} 
where $B_{\phi,\rm{obs}}$ are the values of the updated magnetic field limits, 0.6 and 100 mG, discussed in Sec.~\ref{ssec:obsConstr}  and $B_{\phi,\rm{H_{2}O}}$ is the value of the toroidal component of the magnetic field
for the reference line of our solutions at the position of the H$_2$O maser spot. Then we introduce a third value of $B_{\phi, \rm obs}$ that matches the momentum rate deduced from observational constraints.
Finally, we derive the scaling for the mass density and the pressure from the above as follows:
\begin{equation}
\rho_* = \frac{B_*^2}{4\pi V_*^2}\qquad \rm{and} \qquad P_* = \frac{\mu_{\rm VTST} B_*^2 }{8\pi}.
\end{equation}

\subsection{Integrated quantities}
\label{ssec:intQuant}

As is generally the case for self-similar models, the properties of the jet at a given radius are derived from the reference streamline and extended with the appropriate radial dependence in the form of power law of the parameter $\alpha$ (see Sec.~\ref{sec:eqnMet}, \citep{FerreiraPelletier:1993,VlahakisTsinganos:1998}). 
This parameter is defined as
\begin{equation}
\alpha =  \left(\frac{\varpi_\alpha}{\varpi_\star}\right)^2 = \left(\frac{\varpi(\theta)}{\varpi_\star G(\theta)}\right)^2 
\label{eqn:alpha}
\end{equation}
where $\varpi(\theta)$ is the radial profile of the reference streamline, $G(\theta) = \varpi(\theta)/\varpi_\alpha$, $\varpi_\alpha$ is the cylindrical radius at the AP for the streamline with a given $\alpha$ and $\varpi_\star$ is the scaling length defined in Eq.~\ref{eqn:wscale} using the criterion described in Eq.~\ref{eqn:thetaH2O}.
We give the radial scalings for all the relevant quantities in Appendix~\ref{app:scaling}. 
Using these relations, a given solution can be extended to infinity and towards the polar axis. Expanding a solution over the radial direction is necessary to calculate quantities such as the jet mass loss and the momentum rate which require an integration over a surface perpendicular to the jet axis. Since the geometry of the equations that we adopted has a singularity on the polar axis, for the following calculations of integrated quantities we will consider a \emph{flux tube} defined by inner and outer cylindrical radii, $\varpi_{\rm in}$ and $\varpi_{\rm out}$, or, equivalently, $\alpha_{\rm in}$ and $\alpha_{\rm out}$. 
Once that the scaling length $\varpi_*$ is defined, the inner and outer radii are determined and so are also the streamline labels $\alpha_{\rm in}$ and $\alpha_{\rm out}$, through the equation \ref{eqn:alpha}.

First, we evaluate a density-weighted average velocity for each jet solution at the height of the H$_2$O maser spots over the flux tube area as follows
\begin{equation}
\langle V_{\rm H_2O}\rangle = \frac{  2\pi z_{\rm H_{2}O}^2\int_{\theta_{\rm LRP}}^{\theta_{\rm H_{2}O}} \rho V\frac{\sin(\theta)}{\cos^3(\theta)}d\theta}{ 2\pi z_{\rm H_{2}O}^2 \int_{\theta_{\rm LRP}}^{\theta_{\rm H_{2}O}} \rho\frac{\sin(\theta)}{\cos^3(\theta)}d\theta},
\label{eqn:avVH2O}
\end{equation}
where the relation between $\theta$ and $\alpha$ (or $\varpi$) is defined as
\begin{equation}
\alpha(\theta) = \left(\frac{z_{\rm H_{2}O}\tan(\theta)}{\varpi(\theta)}\right)^2,
\end{equation}
where $\varpi(\theta)$ is calculated on the reference streamline with $\alpha=1$.
We note that, since the velocity decreases with increasing $\alpha$, we expect this average to be dominated by the inner streamlines in the flux tube, while the streamlines close to the outer edge of the flux tube will be slower. For this reason, it is more meaningful to compare the density-averaged velocity with the observed (almost constant) velocity.

The mass loss rate of the jet can be derived as the mass flux flowing from the z=0 surface of the flux tube as follows
\begin{equation}
\dot{M_j} = \int_{\varpi_{\rm in}}^{\varpi_{\rm out}} \rho V \cdot dA.
\end{equation}
The mass loss rate is dependent on the value of the toroidal magnetic field we are considering. Since there is still considerable uncertainty on the strength of the toroidal component of the magnetic fields, we can associate to each solution three mass loss rates corresponding to the minimum and maximum $B_\phi$ in Sec.~\ref{ssec:obsConstr} and the minimum value of $B_\phi$ for which we find matching solutions (see Fig.~\ref{fig:PdotZH2O}). 

Similarly we will give three values for the momentum rate of each jet configurations.
The momentum rate of a jet model is 
\begin{equation}
\dot{P} =  2\pi z_{\rm H_{2}O}^2 \int_{\theta_{\rm LRP}}^{\theta_{\rm H_{2}O}}  \rho V^2 \frac{\sin(\theta)}{\cos^3(\theta)}d\theta,
\end{equation}
where the integration is done over all the streamlines contributing to the flux tube above the H$_2$O maser spot.

\subsection{Comparison results}
\label{ssec:results}

\begin{figure}
\centering
\includegraphics[width=0.99\linewidth]{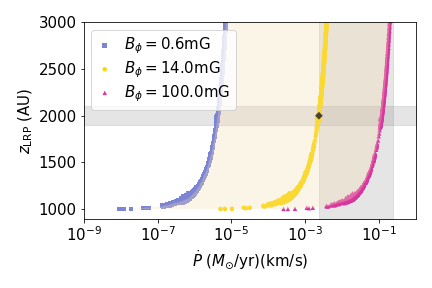}
\caption{Total jet height versus momentum rate for a jet width of 20 AU at H$_2$O for all the solutions in the sample. The blue squares and pink triangles represent the solutions $\dot P$ for a toroidal magnetic field of 0.6 and 100 mG, respectively. The yellow dots are the momentum rates of the solutions with the minimum $B_\phi$ (14 mG) that matches the observed $\dot P$. The light yellow areas highlights the allowed momentum rates within this interval of $B_\phi$. The shaded gray areas show the acceptance intervals for $z_{\rm LRP}$ and $\dot P$.}
\label{fig:PdotZH2O}
\end{figure}

In Fig.~\ref{fig:PdotZH2O} we present the total jet height ($z_{\rm LRP}$) versus the momentum rate of all the solutions in our sample. The black diamond marks the observed $\dot P$ at the observed total jet height.  
The shaded gray horizontal and vertical areas show the intervals for $\dot P$ and $z_{\rm LRP}$ we use to define a solution as a good match.
The shaded light yellow area between the blue squares and the magenta triangles define the values of the momentum rates that are allowed within the range of the toroidal magnetic field derived from observations. We see that for any value of $B_\phi$ the solutions fall on a curve with little-to-none scattering introduced by the variation of the other jet properties. 
We produce this plot once we have set the half-width of the jet, but before introducing the other constraints on velocity and density and we find that a toroidal magnetic field at the H$_2$O maser spots of at least 14 mG is required for the jet solutions to have a comparable or higher momentum rate than the observed one.

Among the solutions found at the interception of the shaded gray areas in Fig.~\ref{fig:PdotZH2O} for the given choice of the jet radius ($\varpi_{\rm H2O}=20$ AU) and toroidal magnetic field ($B_\phi = 14$ mG), we present a sample of 8 jet configurations which satisfy all the constraints on density, velocity, total jet height and momentum rate.
We report the parameters and the relevant scaled quantities in Tab.~\ref{tab:selsol}.
Given the observational constraints (Sec.~\ref{ssec:obsConstr}) and the tight correlation that exist between the jet total extent and the momentum rate, we are left with solutions having the same angular position and collimation angle at the AP, $\theta_{\rm A}=14\degree$ and $\psi_{\rm A}=79\degree$ (and $\psi_0=35\degree$) respectively, for the given choice of the parameter $F$ ($F=0.75$) and the polytropic index of the gas ($\Gamma=5/3$). As a reference, we give the typical BP solution parameters ($k_{\rm BP}= 0.03,~\lambda_{\rm BP}= 30,~\psi_0 = 32\degree$) and we report in Tab.~\ref{tab:selsol} our parameters in BP units. We remind the reader that the equations that we adopted differ from the classical BP because we do not neglect the enthalpy of the gas.
\begin{figure}
\centering
\includegraphics[width=0.99\linewidth]{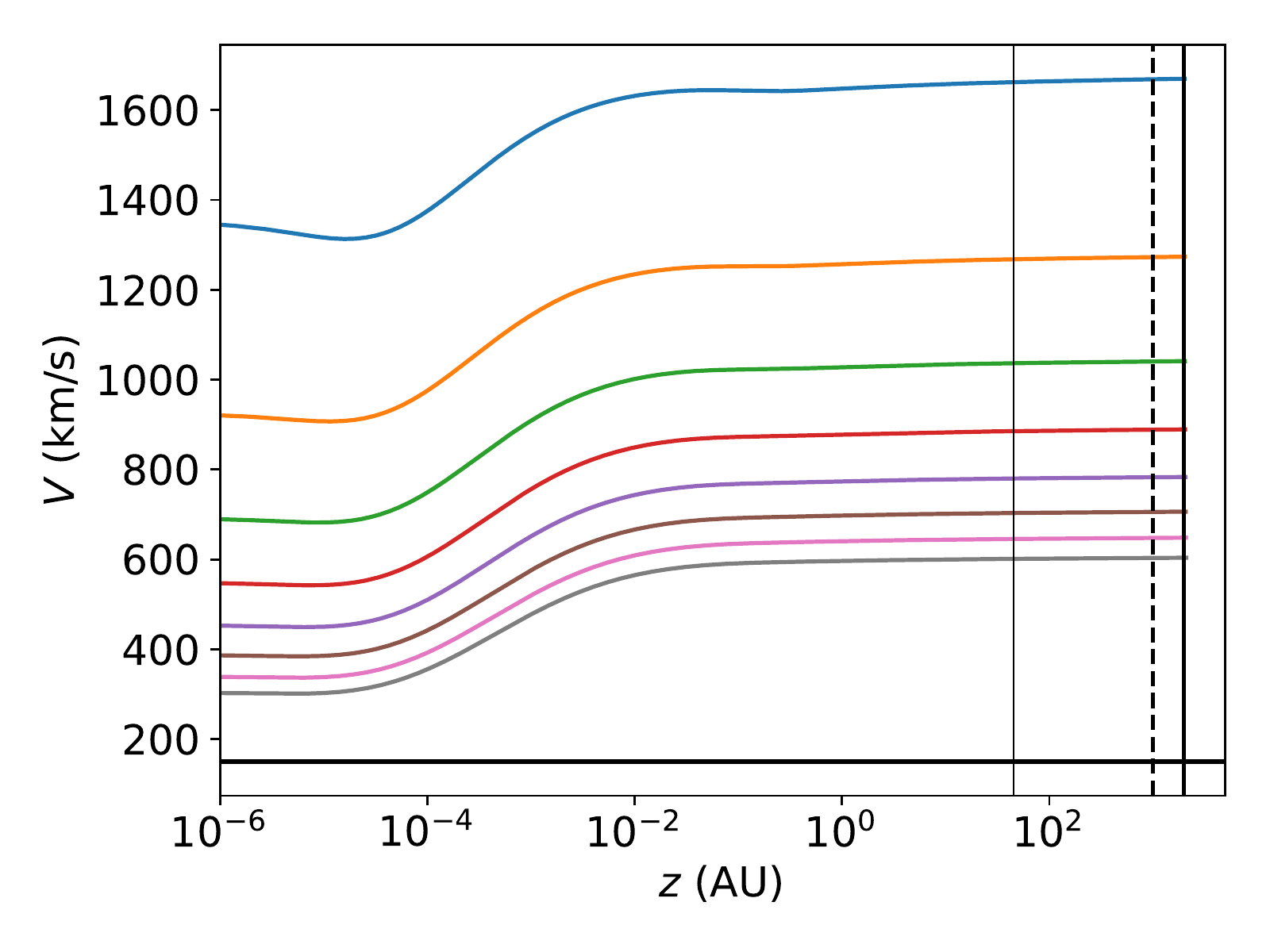}
\caption{Velocity versus jet height of the scaled reference streamlines for $\alpha=1$, which are the outermost streamlines of our selected jet configurations. The solid horizontal line is the observed velocity at the H$_2$O masers. The vertical thin solid line is the "base" of the jet as seen in the CO observations, and the vertical dashed solid line is the location of the H$_2$O maser spot. The thick solid vertical line is at z = 2000 AU.}
\label{fig:VprofileW43A}
\end{figure}

\begin{table*}
\centering
\setlength\tabcolsep{4.5 pt}
\setlength\extrarowheight{5pt}
 \caption{Parameters of the selected solutions as a good match to W43A. All solutions have models parameters $F=0.75$, $\Gamma=5/3$, $\theta_{\rm A}=14\degree$, $\psi_{\rm A}=79\degree$ and scaling parameters $\varpi_{\rm H2O} = 20$ AU, $z_{\rm H2O} = 1000$ AU and $B_\phi = 14$ mG.}
 \begin{tabular}{l | c c c c c c c c}
 \hline
 Model & S1 & S2 & S3 & S4 & S5 & S6 & S7 & S8 \\
 \hline
$k_{\rm VTST}$ & 1.5 & 2.0 & 2.5 & 3.0 & 3.5 & 4.0 & 4.5 & 5.0   \\
$\mu_{\rm VTST}$ & 0.7957  & 1.3122    &   1.9295   & 2.6433    &  3.4504 & 4.3488 & 5.3373 & 6.4149  \\
$\lambda_{\rm VTST}$ & 0.3155 &  0.3185 &   0.3214 &   0.3239 &   0.3261 & 0.3280 & 0.3295 & 0.3307 \\ 
$\epsilon_{\rm VTST}$ & 11.2599 &  11.4833 &   11.7117 &   11.9394 &   12.1626 & 12.3788 & 12.5862 & 12.7840 \\ 
$\theta_{\rm MFP}$ & $2.1866\times 10^{-2}$ &  $2.1753\times 10^{-2}$ &    $2.1623\times 10^{-2}$ &   $2.1480\times 10^{-2}$ &   $2.1327\times 10^{-2}$ & $2.1166\times 10^{-2}$ & $2.1000\times 10^{-2}$ & $2.0830\times 10^{-2}$ \\
$\theta_{\rm MSP}$ & 0.7494 & 0.6046 & 0.5194 & 0.4631 & 0.4232 & 0.3938 & 0.3712 & 0.3535\\
\hline
$\langle V_{\rm H_{2}O} \rangle$ (km/s) & 1982 & 1510 & 1238 & 1059 & 934 & 843 & 775 & 722 \\
$\langle V_{p,{\rm MSP}} \rangle$ (km/s) & 1405 & 1057 & 859 & 731 & 642 & 578 & 530 & 494 \\
$\langle n_{\rm H_{2}O} \rangle$ (cm$^{-3}$) & $8.05\times 10^{5}$ & $1.39\times 10^{6}$ & $2.10\times 10^{6}$ & $2.87\times 10^{6}$ & $3.72\times 10^{6}$ & $4.60\times 10^{6}$ & $5.49\times 10^{6}$ & $6.37\times 10^{6}$\\
$\dot M_j$ (M$_\odot$/yr) & $1.24\times 10^{-6}$ & $1.58\times 10^{-6}$ & $1.95\times 10^{-6}$ & $2.24\times 10^{-6}$ & $2.64\times 10^{-6}$ & $2.87\times 10^{-6}$ & $3.08\times 10^{-6}$ & $3.26\times 10^{-6}$ \\
$\dot P$ (M$_\odot$/yr)(km/s) & $2.38\times 10^{-3}$ & $2.41\times 10^{-3}$ & $2.41\times 10^{-3}$ & $2.42\times 10^{-3}$ & $2.43\times 10^{-3}$ & $2.45\times 10^{-3}$ & $2.48\times 10^{-3}$ & $2.51\times 10^{-3}$ \\
$P$ (g cm/s) & $3.07\times 10^{37}$ & $3.12\times 10^{37}$ & $3.12\times 10^{37}$ & $3.13\times 10^{37}$ & $3.14\times 10^{37}$ & $3.17\times 10^{37}$ & $3.20\times 10^{37}$ & $3.24\times 10^{37}$ \\
\hline
$k_{\rm BP}$ & 0.0032 & 0.0042 & 0.0052 & 0.0062 & 0.0071 & 0.0080 & 0.0088 & 0.0097   \\
$\mu_{\rm BP}$ & 13.1344  & 8.3814    &  5.9173   & 4.4605    &  3.5194 & 2.8716 & 2.4038 & 2.0532  \\
$\lambda_{\rm BP}$ & 1.6307 &  1.2404 &   1.0059 &   0.8491 &   0.7364 & 0.6513 & 0.5844 & 0.5303 \\
$\epsilon_{\rm BP}$ & $8.3257\times 10^{-2}$ &  $4.7334\times 10^{-2}$ &   $3.0601\times 10^{-2}$ &   $2.1451\times 10^{-2}$ &   $1.5895\times 10^{-2}$ & $1.2265\times 10^{-2}$ & $9.7593\times 10^{-3}$ & $7.9553\times 10^{-3}$ \\
    \hline
\end{tabular}
\label{tab:selsol}
\end{table*}

We notice that the only other parameter that has not been constrained is the mass loss parameter $k_{\rm VTST}$. This parameter is proportional to the mass-to-magnetic flux ratio, which leads to an uncertainty on the dimensionless angular momentum, $\lambda_{\rm VTST}$, and the parameter regulating the entropy of the gas, $\mu_{\rm VTST}$. Such spread in the values of the above parameters is reflected in the uncertainties on the average velocities and densities at the maser spot. However, the resulting momentum rate among the selected models is roughly constant and very close to the momentum rate estimated from the observations ($\dot P \sim 0.0024-0.0025$ M$_\odot$/yr km/s). We note that substantially larger $\dot P$ could be achieved by adopting a larger toroidal magnetic field. 
A shorter jet ($<2000$ AU) would still require a larger $B_\phi$, while a taller jet could yield a larger momentum rate for the same choice of $B_\phi$.

In Fig.~\ref{fig:VprofileW43A} we show the velocity profiles of the jet models given in Tab.~\ref{tab:selsol}.
Given the model assumptions given in Sec.~\ref{ssec:modAssumpt}, we selected solutions that have velocity profiles of the reference ($\alpha=1$) outermost line of the flux tube largely exceeding the average observed velocity of 150 km/s (horizontal black line), so that the MHD jet would be able to transfer momentum to and accelerate the molecular cocoon. 
The inner streamlines ($\alpha < 1$) have the same acceleration profile, but higher speeds due to the relations \ref{eqn:Vpol} and \ref{def:Vphi} given in Appendix \ref{app:scaling}.
We notice that the smaller the mass load parameter, $k_{\rm VTST}$, the larger is the speed. The uncertainty left on $k_{\rm VTST}$, and therefore on the velocity of the MHD jet layer, can only be removed with further observations of the core of the molecular jet of W43A.

While there is a moderate acceleration taking place from the MSP to shortly downstream of the Alfv\'en point, in the portion of the jet observed through the emission of CO, i.e. from 45 AU to 2000 AU  (the region between the thin vertical solid line and the thick vertical solid line in Fig.~\ref{fig:VprofileW43A}), the velocity has already reached its maximum and it stays constant up until the jet tip. 
The total velocity is entirely poloidal, while the toroidal component is close to zero along the entire jet extent. Under these circumstances, the magnetic field and the gas are not corotating even upstream of the Alfv\'en surface, which is an indication of a jet driven by thermal pressure (see bottom panels of Fig.~\ref{fig:energyfluxes}-\ref{fig:poloidalF}) as opposed to a magnetically-driven jet (top and middle panels of Fig.~\ref{fig:energyfluxes}-\ref{fig:poloidalF}). Typically these winds are less powerful and they can only achieve higher speeds if a large injection speed ($V_{p,{\rm MSP}}$) is provided (see Tab.~\ref{tab:selsol}). Such high injections speeds are consistent or higher than the initial speeds considered in recent MHD simulations by \citet{Balick:2020}, which are successful in reproducing the qualitative shapes of a sample of pre-planetary nebulae.

In order to make this comparison as complete as possible, we investigate the effect of varying the radius of the jet and of having a main sequence star as the accreting object. Increasing or decreasing the jet radius within the observed range 10-45 AU has the effect of decreasing/increasing the angular position of the AP  and increasing/decreasing its collimation angle. The thinnest jet ($\varpi_{\rm H_{2}O} = 10$ AU, $\theta_{\rm A}=10\degree$, $\psi_{\rm A}=83\degree$) allows two values of the mass loss parameter (1.5 and 2.0) instead of eight, limiting the selection to just two models. The thickest jet ($\varpi_{\rm H_{2}O} = 45$ AU) leaves us solutions with $\theta_{\rm A}$ ($\sim25\degree$) and $\psi_{\rm A}$ ($68\degree$) and excludes $k_{\rm VTST}=1.5$.
We also considered the possibility that the central object may be a main sequence star ($M\sim M_\odot$ and $R_\star\sim R_\odot$), finding our conclusions unaltered, as expected by the mild dependence that our scaling scheme has on the mass of the central star.

\section{Summary}
\label{sec:discussion}

In this paper we discussed the adaptation of the numerical algorithm we presented in Paper I to solve the non-relativistic, radial self-similar MHD equations describing a disk-driven outflow. 
We focused on the study of a large sample of solutions defined by constant Blandford-Payne-like parameter $F$ ($F=0.75$) and polytropic index $\Gamma=5/3$. We recognized similar patterns within the collection of jet configurations that are ultimately ascribed to the cold-to-hot transition that we find recurrently for similar values of the angular position of the Alfv\'en point and the collimation angle at the same position.
We analysed the behaviour of all the relevant jet quantities undergoing this transition and found that:
\begin{itemize}
    \item Cold jets have the largest (dimensionless) angular momentum and they have the lowest enthalpy and plasma-$\beta$ much lower then unity. They are therefore magnetically-dominated jets. They have little-to-none vertical speed upstream of the magnetosonic slow point, but have $|B_\phi/B_p|$ ratios larger than unity. This combination produces twisted streamlines with variable radius, due to the oscillatory behaviour of the transverse forces. The highly wound up magnetic field is also responsible for the relatively high mass load ($\eta \lesssim 1$) of these solutions. At approximately half-way between the Alfv\'en point and the magnetosonic fast point, a large fraction of the magnetic energy has turned into kinetic energy and the jet becomes kinetically-dominated until the last recollimation point. 
    \item Magneto-centrifugal jets are similar to cold jets however the enthalpy is slightly larger, and it plays a role in lifting the gas. These jet models do not suffer oscillations upstream of the magnetosonic slow point. From this point on, these models resemble closely the cold jets. They are, however, the most efficient at accelerating the flow, even though their total energy flux is lower than the purely cold jets.
    \item Hot jets are thermally-dominated jet configurations (plasma-$\beta = P/(B^2/8\pi) \lesssim 1$), where the magnetic field is contributing significantly to the acceleration and collimation of the jet only downstream of the Alfv\'en point. These solutions start off with a large poloidal speed, negative toroidal velocity and $|B_\phi/B_p|<<1$. The acceleration is only mild and they have low energy flux densities. Within this regime, we see two types of energy transfer channels that lead to an increase in kinetic energy. In hot jets the gas pressure is responsible for the acceleration in the initial jet segment, which can extend even just downstream of the Alfv\'en point. A fraction of the thermal energy is transferred to the magnetic energy, which then is used for the last acceleration until the tip of the jet.
\end{itemize}
We then describe a procedure for the identification of specific jet solutions to be compared to an astrophysical source, in our case the water fountain W43A.
W43A is believed to be an asymptotic giant branch star in the process of becoming a planetary nebula. During this current, short-lived phase, the source is launching collimated molecular jets, the nature of which is debated.

We assume that the jets of W43A is launched by a disk-driven ionized inner shell, which is surrounded by a molecular jet sheath.
Since the true nature or even the existence of such a jet core is unknown, we adopted the constraints on the molecular gas as upper/lower limits to the corresponding quantities of the atomic jet, namely the size, hydrogen number density, velocity and magnetic field strength to identify possible jet configuration. 
We conducted an exhaustive examination of our sample and we established that, given the observed molecular properties within the jets of W43A and our (large, but finite) collection of solutions, the most suitable jet model for the jets of W43A is a thermally-dominated jet configuration with a high injection speed, but not efficiently accelerating for most of its extent. We found that the strength of the toroidal component of the magnetic field is the parameter that affects the most this comparison. 
This procedure can be applied to other sources, for which the magnetic field has been difficult to measure, in order to determine a range of plausible magnetic field strengths given observational constraints on density, velocity and jet size.
In future works, we will expand our grid of solutions to the additional two dimensions, namely the radial scaling of the current, $F$, and the polytropic index, $\Gamma$, and compare the full sample to other astrophysical sources.

\section*{Acknowledgements}

CC and WV acknowledge support from the Swedish Research Council (VR).

\section*{Data Availability}
A catalogue of all the solutions that have been found is available on request to the main author.



\bibliographystyle{mnras}
\bibliography{mybiblio} 

\onecolumn
\appendix

\section{Coefficients of the Bernoulli and transverse equations}
\label{app:numsden}

In eq.~\ref{eqn:detform}, we gave a general form to which both the Bernoulli and the transfield equations can be reduced. Here we provide the explicit form of the coefficients $A_i, B_i, C_i$ with $i=1,2$ for both equations as follows

\begin{align}
     & A_1  = \frac{\cos^2(\psi+\theta)}{\sin^2(\theta)}\left[-2\lambda_{\rm VTST}^2 \frac{M^2}{G^2} \frac{(1-G^2)^2}{(1-M^2)^3}+\Gamma\frac{\mu_{\rm VTST}}{M^{2\Gamma}}\right] - 2\frac{M^2}{G^4}\\
     & B_1 =  -2\frac{M^4}{G^4}\tan(\psi+\theta)\\
     & C_1 =   2k_{\rm VTST}^2\frac{\sin(\theta+\psi)\cos(\theta+\psi)}{G\sin(\theta)} - 2\frac{M^4}{G^4}\frac{\cos(\psi)}{\sin(\theta)\cos(\theta+\psi)} + 2\lambda_{\rm VTST}^2\frac{G^2(2M^2-1)-M^4}{(1-M^2)^2}\frac{\cos(\psi)\cos(\psi+\theta)}{\sin^3(\theta)}\\
     & A_2  =  \sin(\theta+\psi)\cos(\theta+\psi)\left[\lambda_{\rm VTST}^2G^2\frac{(1-G^2)^2}{(1-M^2)^3}
         -\Gamma\mu_{\rm VTST}\frac{G^4}{2M^{2(\Gamma+1)}} \right]\\
     & B_2 =    \sin^2(\theta)\left[\frac{1}{\cos^2(\theta+\psi)} - M^2\right]\\
     & C_2 =   \lambda_{\rm VTST}^2\frac{G^2}{M^2}\left(\frac{G^2-M^2}{1-M^2}\right)^2\frac{\sin(\psi)\cos(\psi+\theta)}{\sin(\theta)} +\frac{\sin^2(\theta)}{\cos^2(\psi+\theta)}\left[F-2+\frac{\cos(\psi)\sin(\theta+\psi)}{\sin(\theta)}\right] \nonumber\\
      & \qquad+\lambda_{\rm VTST}^2G^2\frac{(1-G^2)^2}{(1-M^2)^2}\left[F-1+2\frac{G^2}{(1-G^2)}\frac{\cos(\psi)\sin(\psi+\theta)}{\sin(\theta)}\right]
      +\mu_{\rm VTST}(F-2)\frac{G^4}{M^{2\Gamma}}  +k_{\rm VTST}^2\frac{G^3}{M^2}\sin(\theta)\cos^2(\psi+\theta)
\end{align}

\section{Physical scaling}
\label{app:scaling}

We report here the definitions of the density and the velocity and magnetic field components from \citetalias{VTST:2000}. 

\begin{align}
\rho & = \frac{\rho_\star}{M^2}\alpha^{F-3/2}, \quad P = \frac{P_\star}{M^{2\Gamma}}\alpha^{F-2}, \quad {\rm with}\quad \alpha =  \left(\frac{\varpi_\alpha}{\varpi_\star}\right)^2 = \left(\frac{\varpi}{\varpi_\star G}\right)^2 \label{def:densAndP}\\
\boldsymbol{B_p} & = - \frac{B_\star\alpha^{(F-2)/2}}{G^2}\frac{\sin(\theta)}{\cos(\theta+\psi)}\hat{\boldsymbol{b}} = B_r \hat{\boldsymbol{r}} + B_\theta\hat{\boldsymbol{\theta}} = B_p\sin(\theta+\psi)\hat{\boldsymbol{r}}+B_p\cos(\theta+\psi)\hat{\boldsymbol{\theta}},\\
B_p &= - \frac{B_\star\alpha^{(F-2)/2}}{G^2}\frac{\sin(\theta)}{\cos(\theta+\psi)}, \\ 
\boldsymbol{B_\phi}  & = - \lambda_{\rm VTST}\frac{B_\star\alpha^{(F-2)/2}}{G}\frac{(1-G^2)}{(1-M^2)} \hat{\boldsymbol{\phi}}= B_\phi\hat{\boldsymbol{\phi}}\\
\boldsymbol{V_p} & = - \frac{V_\star\alpha^{-1/4}M^2}{G^2}\frac{\sin(\theta)}{\cos(\theta+\psi)}\hat{\boldsymbol{b}} = V_r \hat{\boldsymbol{r}} + V_\theta\hat{\boldsymbol{\theta}} = V_p\sin(\theta+\psi)\hat{\boldsymbol{r}}+V_p\cos(\theta+\psi)\hat{\boldsymbol{\theta}},\label{eqn:Vpol}\\
V_p & = - \frac{V_\star\alpha^{-1/4}M^2}{G^2}\frac{\sin(\theta)}{\cos(\theta+\psi)},  \label{def:Vp} \\
\boldsymbol{V_\phi} & =  \lambda_{\rm VTST}\frac{V_\star\alpha^{-1/4}}{G}\frac{(G^2-M^2)}{(1-M^2)} \hat{\boldsymbol{\phi}} = V_\phi\hat{\boldsymbol{\phi}} \label{def:Vphi} \\
B_\star & = \sqrt{4\pi \rho_\star}V_\star; \quad \quad P_\star = \mu_{\rm VTST} \frac{B_\star^2}{8\pi};  \quad V_\star^2 = \frac{\mathcal{GM}}{\varpi_\star k_{\rm VTST}^2}  \label{def:starred}
\end{align}

If we want to compare our solutions to a real jets, we need to determine the scaling of the physical quantities for a given streamline. Following  \citetalias{VTST:2000}, we give here general scaling relations similar to the ones given in their section 5.1, which are tailored for a subclass of solutions ($M_0 \simeq 0.01, \, G_0 \simeq 0.1$ and F = 0.75). Using the definition of $\alpha$ in \ref{def:densAndP} and of $V_\star$ in \ref{def:starred}, the Keplerian velocity at the footpoint of a streamline (z = 0) can be written as
\begin{equation}
V_{\rm kep,0}  = \sqrt{\frac{\mathcal{GM}}{\varpi_0}} = k V_\star G_0^{-1/2} \alpha^{-1/4} \label{def:Vk0}.
\end{equation}
The sound speed for a polytropic gas is defined as
\begin{equation}
C_{\rm s}^2 =  \frac{d P}{d \rho} = \frac{\Gamma P}{\rho}
\end{equation}
which can be calculated at z=0 and recast by making use of the Eqs. \ref{def:densAndP} and Eqs.\ref{def:starred} as follows
 \begin{equation}
C_{\rm s,0}^2 =  \frac{\mu_{\rm VTST}}{2}  \Gamma V_\star^2\alpha^{-1/2} M_0^{2(1-\Gamma)}.
\end{equation}
The vertical component of the velocity at z=0 is $V_0 \equiv V_z(z=0) =  V_{p,0} \sin\psi_0$, or
\begin{equation}
V_0 \equiv V_z (z=0) = \alpha^{-1/4}M_0^2 G_0^{-2} V_\star
\end{equation}
With these quantities, we can calculate the following ratios
\begin{align}
\left(\frac{C_{\rm s}}{V_0}\right)_0 & = \sqrt{\frac{\mu_{\rm VTST} \Gamma }{2}}  M_0^{-(1+\Gamma)}G_0^{2} \label{def:CsV0}\\
\left(\frac{V_{\rm kep}}{V_0}\right)_0 & = k_{\rm VTST}\, G_0^{3/2} M_0^{-2} \label{VkV0} \\
\left(\frac{C_{\rm s}}{V_{\rm kep}}\right)_0 & =  \sqrt{\frac{\mu_{\rm VTST} \Gamma }{2 k_{\rm VTST}^2}} M_0^{(1-\Gamma)}  G_0^{1/2} 
\end{align}
and, as well, the reference length to scale the solutions to real objects
\begin{align}
\varpi_0=\frac{\mathcal{GM}}{V_{\rm kep}^2} = \frac{\mathcal{GM}}{\left(\frac{V_{\rm kep}}{V_0}\right)^2 V_0^2} \simeq 1.91\times10^5 k_{\rm VTST}^{-2} M_0^4 G_0^{-3} \left(\frac{\mathcal{M}}{\mathcal{M}_\odot}\right) \left(\frac{V_0}{\rm{km\, s^{-1}}}\right)^{-2} R_\odot
\end{align}

\section{Poloidal and transfield forces acting on a fieldline}
\label{app:forces}

In the non -relativistic case it is customary to divide the force acting on the field line in 4 contributions: the kinetic force, the thermal pressure force, the electromagnetic force and the gravitational force. The net force is zero, so we have the following equation:
\begin{equation}
\mathcal{F}_{\rm K} + \mathcal{F}_{\rm T} + \mathcal{F}_{\rm EM} + \mathcal{F}_{\rm G} \equiv \rho(\boldsymbol{V} \cdot\nabla)\boldsymbol{V} + \nabla P -\frac{(\nabla\times \boldsymbol{B})\times \boldsymbol{B}}{4\pi} - \rho\nabla\frac{\mathcal{GM}}{r} = 0. \label{eqn:totForces}
\end{equation}

By taking the inner product with $\hat{\boldsymbol{n}}$, we obtain the projection of the forces in the direction perpendicular to the field line, whereas with the inner product with $\hat{\boldsymbol{b}}$, we obtain the forces along the poloidal direction.

\subsection{Transfield forces}
\label{app:transForce}

The transfield kinetic force:
\begin{align}
\mathcal{F}_{\rm K, \perp}  \equiv \rho(\boldsymbol{V} \cdot\nabla)\boldsymbol{V} \cdot \hat{\boldsymbol{n}} =  &\,  \rho\cos(\theta+\psi)\left[V_r\frac{\partial V_r}{\partial r} + \frac{V_\theta}{r}\frac{\partial V_r}{\partial \theta} - \frac{V_\theta^2+V_\phi^2}{r}\right] 
 -\rho \sin(\theta+\psi)\left[V_r\frac{\partial V_\theta}{\partial r} + \frac{V_\theta}{r}\frac{\partial V_\theta}{\partial \theta} + \frac{V_\theta V_r}{r} - \frac{V_\phi^2}{r\tan(\theta)}\right] \nonumber\\
= &\, \rho\frac{V_p^2}{\varpi}\sin(\theta)\cos(\theta+\psi)\frac{d\psi}{d\theta} +\rho \frac{V_\phi^2}{\varpi}\sin(\psi)\nonumber\\
= &\, \rho \left(- \frac{V_\star\alpha^{-1/4}M^2}{G^2}\frac{\sin(\theta)}{\cos(\theta+\psi)}\right)^2\sin(\theta)\frac{\cos(\theta+\psi)}{\varpi}\frac{d\psi}{d\theta} + \rho \left( \lambda_{\rm VTST}\frac{V_\star\alpha^{-1/4}}{G}\frac{(G^2-M^2)}{(1-M^2)}\right)^2\frac{\sin(\psi)}{\varpi}\nonumber\\
= & \left[\frac{B_\star^2\alpha^{F-2}}{4\pi \varpi G^4}\frac{\sin(\theta)}{\cos(\theta+\psi)}\right] \left\{M^2\sin^2(\theta)\frac{d\psi}{d\theta}+\lambda_{\rm VTST}^2\frac{G^2}{M^2}\frac{(G^2-M^2)^2}{(1-M^2)^2}\frac{\sin(\psi)}{\sin(\theta)}\cos(\theta+\psi)\right\}
\end{align}
The transfield thermal pressure force:
\begin{align}
\mathcal{F}_{\rm T, \perp}  \equiv\nabla P \cdot \hat{\boldsymbol{n}} = &  \left\{\frac{\partial P}{\partial r} \hat{\boldsymbol{r}}  + \frac1r \frac{\partial P}{\partial \theta} \hat{\boldsymbol{\theta}}\right\} \cdot \hat{\boldsymbol{n}} \nonumber\\
= &\, 2P \frac{\sin(\theta)}{\varpi \cos(\theta+\psi)}\left((F-2)(\cos^2(\theta+\psi)+\sin^2(\theta+\psi))+\frac{\Gamma}{2M^2}\sin(\theta+\psi) \cos(\theta+\psi)\frac{dM^2}{d\theta}\right)\nonumber\\
= &  \left[\frac{B_\star^2\alpha^{F-2}}{4\pi \varpi G^4}\frac{\sin(\theta)}{\cos(\theta+\psi)}\right] \left(\frac{G^4\mu_{\rm VTST}}{M^{2\Gamma}}\right)\left\{(F-2)+\frac{\Gamma}{2M^2}\sin(\theta+\psi) \cos(\theta+\psi)\frac{dM^2}{d\theta}\right\}.
\end{align}
The transfield electromagnetic force:
\begin{align}
\mathcal{F}_{\rm EM, \perp}  \equiv - \frac{(\nabla\times \boldsymbol{B})\times \boldsymbol{B}}{4\pi} \cdot \hat{\boldsymbol{n}} = & - \frac{1}{4\pi} \left[\frac{B_\phi}{r}\left(\frac{1}{\sin(\theta)}\frac{\partial B_r}{\partial \phi}-\frac{\partial (rB_\phi)}{\partial r}\right) - \frac{B_\theta}{r}\left(\frac{\partial (rB_\theta)}{\partial r} -\frac{\partial B_r}{\partial \theta}\right)\right] \hat{\boldsymbol{r}}\cdot \hat{\boldsymbol{n}}\nonumber\\
& -\frac{1}{4\pi} \left[ \frac{B_r}{r}\left(\frac{\partial (rB_\theta)}{\partial r} -\frac{\partial B_r}{\partial \theta}\right) - \frac{B_\phi}{r\sin(\theta)}\left(\frac{\partial (\sin(\theta)B_\phi)}{\partial \theta} - \frac{\partial B_\theta}{\partial \phi}   \right)\right] \hat{\boldsymbol{\theta}}\cdot \hat{\boldsymbol{n}}\nonumber\\
= & \frac{B_\phi^2}{4\pi\varpi}\frac{\sin(\theta)}{\cos(\theta+\psi)}\left\{(F-1)+ \frac{2\cos(\psi)\sin(\theta+\psi)}{\sin(\theta)}\frac{G^2}{1-G^2} - \frac{\sin(\theta+\psi)\cos(\theta+\psi) }{(1-M^2)}\frac{dM^2}{d\theta}\right\}\nonumber\\
& +\frac{B_p^2}{4\pi\varpi}\frac{\sin(\theta)}{\cos(\theta+\psi)}\left\{(F-2)+\frac{\cos(\psi)\sin(\theta+\psi)}{\sin(\theta)}-\frac{d\psi}{d\theta} \right\}\nonumber\\
= & \left[\frac{B_\star^2\alpha^{F-2}}{4\pi \varpi G^4}\frac{\sin(\theta)}{\cos(\theta+\psi)}\right]\lambda_{\rm VTST}^2\left(\frac{G(1-G^2)}{(1-M^2)}\right)^2\left\{(F-1)+ \frac{2\cos(\psi)\sin(\theta+\psi)}{\sin(\theta)}\frac{G^2}{1-G^2} - \frac{\sin(\theta+\psi)\cos(\theta+\psi) }{(1-M^2)}\frac{dM^2}{d\theta}\right\}\nonumber\\
 & +\left[\frac{B_\star^2\alpha^{F-2}}{4\pi \varpi G^4}\frac{\sin(\theta)}{\cos(\theta+\psi)}\right] \frac{\sin^2(\theta)}{\cos^2(\theta+\psi)}\left\{(F-2)+\frac{\cos(\psi)\sin(\theta+\psi)}{\sin(\theta)}-\frac{d\psi}{d\theta} \right\}
\end{align}
The transfield gravitational force:
\begin{align}
\mathcal{F}_{\rm G, \perp}  \equiv -\rho\nabla\frac{\mathcal{GM}}{r} \cdot \hat{\boldsymbol{n}} = & -  \frac{\rho_\star}{M^2}\alpha^{F-3/2} \mathcal{GM}\frac{\partial }{\partial r}\left(\frac1r\right) \hat{\boldsymbol{r}}\cdot \hat{\boldsymbol{n}}\nonumber\\
 = & \frac{\rho_\star}{M^2}\alpha^{F-3/2} \mathcal{GM}\frac{\sin^2(\theta)}{\varpi^2}\cos(\theta+\psi)\nonumber\\
 = &  \left[\frac{B_\star^2\alpha^{F-2}}{4\pi \varpi G^4}\frac{\sin(\theta)}{\cos(\theta+\psi)}\right] \left(\frac{k_{\rm VTST}^2\sin(\theta)}{G}\right)\frac{G^4}{M^2}\cos^2(\theta+\psi)
\end{align}

\subsection{Poloidal forces}
\label{app:polForce}

In this section, we give the forces along the poloidal directions.\\
The poloidal kinetic force:
\begin{align}
\mathcal{F}_{\rm K, \parallel}  \equiv \rho(\boldsymbol{V} \cdot\nabla)\boldsymbol{V} \cdot \hat{\boldsymbol{b}} & = \rho\sin(\theta+\psi)\left[V_r\frac{\partial V_r}{\partial r} + \frac{V_\theta}{r}\frac{\partial V_r}{\partial \theta} - \frac{V_\theta^2+V_\phi^2}{r}\right] 
+\rho \cos(\theta+\psi)\left[V_r\frac{\partial V_\theta}{\partial r} + \frac{V_\theta}{r}\frac{\partial V_\theta}{\partial \theta} + \frac{V_\theta V_r}{r} - \frac{V_\phi^2}{r\tan(\theta)}\right] \nonumber\\
&  =  \rho\frac{V_p^2}{r}\frac{\cos(\theta+\psi)}{M^2} \frac{dM^2}{d\theta} +\rho  \frac{V_p^2}{r}\sin(\theta+\psi)\frac{d\psi}{d\theta} -\rho \frac{V_p^2 + V_\phi^2}{r} \frac{\cos(\psi)}{\sin(\theta)}\nonumber\\
& =  \left(\frac{B_\star^2}{4\pi V_\star^2}\frac{\alpha^{F-3/2}}{M^2}\right)\frac{\sin(\theta)}{\varpi} \left\{\left(- \frac{V_\star\alpha^{-1/4}M^2}{G^2}\frac{\sin(\theta)}{\cos(\theta+\psi)}\right)^2 \left[\frac{\cos(\theta+\psi)}{M^2} \frac{dM^2}{d\theta} +\sin(\theta+\psi)\frac{d\psi}{d\theta} -\frac{\cos(\psi)}{\sin(\theta)}\right] \right.\nonumber\\ 
& \qquad \qquad \qquad \qquad  \qquad \qquad \left.   -\left(\lambda_{\rm VTST}\frac{V_\star\alpha^{-1/4}}{G}\frac{(G^2-M^2)}{(1-M^2)}\right)^2 \frac{\cos(\psi)}{\sin(\theta)}\right\}\nonumber\\
& = \left[\frac{B_\star^2\alpha^{F-2}}{4\pi \varpi G^4}\frac{\sin(\theta)}{\cos(\theta+\psi)}\right]\left\{\sin^2(\theta)\frac{dM^2}{d\theta} + M^2\sin^2(\theta)\tan(\theta+\psi)\frac{d\psi}{d\theta} - M^2\frac{\sin(\theta)\cos(\psi)}{\cos(\theta+\psi)} \right.\nonumber\\
& \qquad \qquad \qquad \qquad \qquad \qquad \left.- \lambda_{\rm VTST}^2\frac{G^2}{M^2}\frac{(G^2 -M^2)^2}{(1-M^2)^2}\frac{\cos(\psi)\cos(\theta+\psi)}{\sin(\theta)}\right\}.
\end{align}
The thermal pressure force:
\begin{align}
\mathcal{F}_{\rm T, \parallel}  \equiv \nabla P \cdot \hat{\boldsymbol{b}} = &  \left\{\frac{\partial P}{\partial r} \hat{\boldsymbol{r}}  + \frac1r \frac{\partial P}{\partial \theta} \hat{\boldsymbol{\theta}}\right\} \cdot \hat{\boldsymbol{b}} \nonumber\\
= & - P \frac{\sin(\theta)}{\varpi \cos(\theta+\psi)}\left(\cos^2(\theta+\psi)\frac{\Gamma}{M^2}\frac{dM^2}{d\theta}\right)\nonumber\\
= & - \left[\frac{B_\star^2\alpha^{F-2}}{4\pi \varpi G^4}\frac{\sin(\theta)}{\cos(\theta+\psi)}\right] \left(\frac{G^4\mu_{\rm VTST}}{M^{2\Gamma}}\right)\cos^2(\theta+\psi)\frac{\Gamma}{2M^2}\frac{dM^2}{d\theta}.
\end{align}
The electromagnetic force:
\begin{align}
\mathcal{F}_{\rm EM, \parallel}  \equiv - \frac{(\nabla\times \boldsymbol{B})\times \boldsymbol{B}}{4\pi} \cdot \hat{\boldsymbol{b}} = & - \frac{1}{4\pi} \left[\frac{B_\phi}{r}\left(\frac{1}{\sin(\theta)}\frac{\partial B_r}{\partial \phi}-\frac{\partial (rB_\phi)}{\partial r}\right) - \frac{B_\theta}{r}\left(\frac{\partial (rB_\theta)}{\partial r} -\frac{\partial B_r}{\partial \theta}\right)\right] \hat{\boldsymbol{r}}\cdot \hat{\boldsymbol{b}}\nonumber\\
& - \frac{1}{4\pi} \left[ \frac{B_r}{r}\left(\frac{\partial (rB_\theta)}{\partial r} -\frac{\partial B_r}{\partial \theta}\right) - \frac{B_\phi}{r\sin(\theta)}\left(\frac{\partial (\sin(\theta)B_\phi)}{\partial \theta} - \frac{\partial B_\theta}{\partial \phi}   \right)\right] \hat{\boldsymbol{\theta}}\cdot \hat{\boldsymbol{b}}\nonumber\\
= & - \frac{1}{4\pi}\frac{B_\phi^2}{r}\left[\frac{2G^2}{1-G^2}\frac{\cos(\psi)}{\sin(\theta)} - \frac{\cos(\theta + \psi)}{1-M^2}\frac{dM^2}{d\theta}\right]\nonumber\\
= & - \left[\frac{B_\star^2\alpha^{F-2}}{4\pi \varpi G^4}\frac{\sin(\theta)}{\cos(\theta+\psi)}\right] \lambda_{\rm VTST}^2\frac{G^2(1-G^2)^2}{(1-M^2)^2}\left[\frac{2G^2}{1-G^2}\frac{\cos(\psi)\cos(\theta+\psi)}{\sin(\theta)} - \frac{\cos^2(\theta + \psi)}{1-M^2}\frac{dM^2}{d\theta}\right]
\end{align}
The gravitational force:
\begin{align}
\mathcal{F}_{\rm G, \parallel}  \equiv - \rho\nabla\frac{\mathcal{GM}}{r} \cdot \hat{\boldsymbol{b}} = & - \frac{\rho_\star}{M^2}\alpha^{F-3/2} \mathcal{GM}\frac{\partial }{\partial r}\left(\frac1r\right) \hat{\boldsymbol{r}}\cdot \hat{\boldsymbol{b}}\nonumber\\
 = & \frac{\rho_\star}{M^2}\alpha^{F-3/2} \mathcal{GM}\frac{\sin^2(\theta)}{\varpi^2}\sin(\theta+\psi)\nonumber\\
 = & \left[\frac{B_\star^2\alpha^{F-2}}{4\pi \varpi G^4}\frac{\sin(\theta)}{\cos(\theta+\psi)}\right] \left(\frac{k_{\rm VTST}^2\sin(\theta)}{G}\right)\frac{G^4}{M^2}\sin(\theta+\psi)\cos(\theta+\psi)
\end{align}






\bsp	
\label{lastpage}
\end{document}